\documentclass[11pt]{article}
\usepackage{amsfonts,amsmath,amssymb,epsfig,footmisc,here,latexsym,setspace,verbatim} 
\usepackage[all]{xy}
\usepackage{hyperref,color}
\usepackage{mathabx}

\newcommand{\BlackBoxes}{\global\overfullrule5pt}

\BlackBoxes

\def\varepsilon{{\varepsilon}}

\newcommand*\diff{\mathop{}\!\mathrm{d}}

\newcommand{\be}{\begin{equation}}
\newcommand{\ee}{\end{equation}}
\newcommand{\bea}{\begin{eqnarray}}
\newcommand{\eea}{\end{eqnarray}}
\newcommand{\beas}{\begin{eqnarray*}}
\newcommand{\eeas}{\end{eqnarray*}}

\newtheorem{theorem}{Theorem}[section]
\newtheorem{definition}[theorem]{Definition}
\newtheorem{proposition}[theorem]{Proposition}
\newtheorem{corollary}[theorem]{Corollary}
\newtheorem{lemma}[theorem]{Lemma}
\newtheorem{remark}[theorem]{Remark}
\newtheorem{example}[theorem]{Example}
\newtheorem{examples}[theorem]{Examples}

\newtheorem{foo}[theorem]{Remarks}

\newenvironment{proof}{\addvspace{\medskipamount}\par\noindent{\it Proof}\, }
{\unskip\nobreak\hfill$\Box$\par\addvspace{\medskipamount}}

\newcommand{\E}[1]{{\rm E}\left[#1\right]}

\newcommand{\EQ}[1]{{\rm E}_{Q}\left[#1\right]}

\newcommand{\var}[1]{{\rm Var}\left(#1\right)}

\newcommand{\N}{\mathbb{N}} 
\newcommand{\Q}{{\mathbb Q}}

\newcommand{\R}{{\mathbb R}}

\newcommand{\Mcal}{{\mathcal M}}
\newcommand{\Ucal}{{\mathcal U}}

\newcommand{\Fcal}{{\mathcal F}}
\newcommand{\Qcal}{{\mathcal Q}}
\newcommand{\avar}{\operatorname{AV@R}}
\DeclareMathOperator{\essinf}{ess\,inf}

\DeclareMathOperator{\dom}{dom}

\def\F{\mathcal{F}}

\begin{document}
\title{Asymptotic Analysis of Risk Premia
Under Linear Risk Sharing with Law-Invariant Risk Measures\thanks{%: Non Olet Ad Infinitum
%%This
We are very grateful to
Hans F\"ollmer, Frank Riedel, Mitja Stadje and participants at the 2024 workshop on the Foundations and Applications of Decentralized Risk Sharing (FADeRiS) in Ulm 
for comments and suggestions.
This research was funded in part by the Netherlands Organization for
Scientific Research (Laeven) under grants NWO VIDI-2009 and VICI-2019/20.}
\vskip 1cm}
\author{Thomas Knispel\\
{\footnotesize Department of Business and Economics}\\
{\footnotesize Berlin School of Economics and Law}\\
{\footnotesize {\tt thomas.knispel@hwr-berlin.de}}\\
\and Roger J.~A. Laeven\thanks{Corresponding author.}\\
{\footnotesize Department of Quantitative Economics}\\
{\footnotesize University of Amsterdam, CentER}\\
{\footnotesize and EURANDOM}\\
{\footnotesize {\tt R.J.A.Laeven@uva.nl}}\\
%{\footnotesize $^{\sharp}$Corresponding author}\\
\and Gregor Svindland\\
{\footnotesize Institute of Actuarial and Financial Mathematics and House of Insurance}\\ {\footnotesize Leibniz University Hannover}\\
{\footnotesize {\tt gregor.svindland@insurance.uni-hannover.de}}\\[1cm]}
\date{June 23, 2026}
\maketitle
\begin{abstract}
We investigate the asymptotic behavior of the risk premium associated with a linear risk sharing contract in an infinitely expanding risk pool. 
We consider general preferences represented by law-invariant robust utility functionals. 
These preferences encompass the rank-dependent utility model as a special case. 
We also examine Pareto optimality of general and, in particular, linear risk sharing rules with these preferences. 
Our analysis is not limited to the classical i.i.d.\ setting, but allows for heterogeneous risks. 
Two case studies on actuarial pricing for independent but heterogeneous risks illustrate our results.\\[4mm]\noindent\textbf{Keywords:} Risk premium; Risk sharing; Pareto optimality; Large risk pools; Heterogeneous losses; %Ambiguity; Robust preferences;
Law invariance; Probabilistic sophistication;
%Comonotonicity;
Rank-dependent utility. %; Exponential utility; Relative entropy.
%Variational and homothetic preferences;
%Robustness;
%Expanding pools;
%Esscher transform;
%Translation invariance;
%Convexity.
\\[4mm]\noindent\textbf{AMS 2010 Classification:} Primary: 91B06, 91B16, 91B30; Secondary: 60E15,
62P05.
%%60E15: inequalities, stochastic ordering
%%62P05: application to actuarial science or financial mathematics
%%91B06: Decision theory
%%91B16: Utility theory
%%91B30: Risk theory, insurance
\\[4mm]\noindent\textbf{JEL Classification:} D81, G10, G20.
\end{abstract}

\makeatletter
\makeatother
\maketitle

\onehalfspacing

%\doublespacing

\newpage

\section{Introduction}

Sharing of risk, or risk exchange by redistributing risk among economic agents, %in a pool
is at the heart of economics, insurance and finance.
Its potential benefits and welfare implications have been analyzed in a large literature
that starts with
Borch \cite{B60,B62}
Arrow \cite{A63}, Wilson \cite{W68}
and DuMouchel \cite{D68};
see also the early
Arrow \cite{A53}.
As the benefits of risk sharing often grow
with the multiplicity $n$ of risks,
there are clear incentives for the formation of large pools of risk.
For such pools, by the %(Strong)
Law of Large Numbers (whenever valid),
the average risk is close %`close'
to its expected value.
That is, by redistributing and subdividing risks in a sufficiently large pool,
a nearly riskless situation can be established.\footnote{See Samuelson \cite{S63} for an insightful perspective.} 

In this paper, we are concerned with the speed at which the benefits of risk sharing grow when forming large risk pools.
The convergence of the average risk---i.e., the canonical equal risk sharing contract in homogeneous risk pools---to its expected value already follows from the Law of Large Numbers,
with corresponding convergence rate $n$.\footnote{See Prop.~3.1 in Denuit and Robert \cite{DR21} for an extension of this result to linear risk sharing rules.}
We are interested in the \emph{economic evaluation} of the risk sharing contract, viewed through the lens of the risk premium, rather than the convergence of the risk sharing contract itself.
The limiting behavior of the risk premium is, however, more delicate.

This paper explicitly derives the limiting behavior of the risk premium
associated with linearly shared risks in an infinitely expanding pool of risks.
We assume general preferences represented by law-invariant robust utility functionals, encompassing rank-dependent utility preferences.
Our analysis allows for heterogeneous initial loss distributions, as long as the sequence of risks satisfies independence, a Law of Large Numbers and a Central Limit Theorem. 

We establish that the convergence rate, i.e., the `speed' at which the benefits of risk sharing grow with the multiplicity of risks, can be of order $n$
or only $n^{1/2}$, depending upon the agent's precise preferences determining the risk premium.
We prove that a rate of order $n$ essentially necessitates expected utility preferences, whereas genuinely robust preferences lead to a slower convergence rate of order $n^{1/2}$.
%In the former case, a Law of Large Numbers suffices, in the latter case a Central Limit Theorem is required
%to establish formal convergence proofs.
As a by-product, we show that this dichotomy can be linked to first- and second-order risk aversion (Segal and Spivak \cite{SS90}) and examine a measure of first-order risk aversion that occurs naturally in our analysis.

We focus attention on linear or, more generally, affine risk sharing rules. 
This type of risk sharing is widely assumed in the literature and in practice, see Huang and Litzenberger \cite{HL85} and Pratt \cite{P00}, because the corresponding risk sharing rules are easy to communicate to different stakeholders, can readily be implemented, and can be justified under standard expected utility preferences. 
However, an open question is whether this type of risk sharing is optimal under general preferences. 
We give a partial answer to this question by studying Pareto optimality of general and, in particular, linear risk sharing rules for the large class of  preferences under consideration. 
This is based on a generalization of Borch's (\cite{B60}) theorem that we provide. 
Given the relevance of Borch's theorem as well as the popularity of linear risk sharing, this part of the paper may be of independent interest.  

This paper fits to the rapidly growing literature on the problem of risk sharing under general preferences;
see e.g.,
Carlier and Dana \cite{CD03},
Heath and Ku \cite{HK04},
Barrieu and El Karoui \cite{BeK05,BeK09},
Dana and Scarsini \cite{DS07},
Jouini, Schachermayer and Touzi \cite{JST08},
%Kiesel and R\"uschendorf \cite{KR08},
%Ludkovski and R\"uschendorf \cite{LR08},
Filipovi\'c and Svindland \cite{FS08},
Dana \cite{D11},
F\"ollmer and Knispel \cite{FK12},
Laeven and Stadje \cite{LS13},
Ravanelli and Svindland \cite{RS14}, and the references therein.
To our best knowledge, except in trivial cases,
these papers do not establish the asymptotic behavior of the associated risk premia.
In \cite{KLS16}, we analyze the counterpart of this problem for homogeneous i.i.d.\ risks, hence simple equal ($1/n$) risk sharing, with standard preferences that are restricted to be linear in probabilities.

The outline of this paper is as follows. 
In Section~\ref{sec:pre}, we introduce the model and in particular the class of preferences we are considering. 
Section~\ref{sec:main} contains the main result on the limiting behavior of the corresponding risk premium. 
Section~\ref{sec:Pareto} is devoted to the question of Pareto optimality of general and, in particular, linear risk sharing rules. 
Applications in actuarial pricing for large portfolios with independent but heterogeneous risks are presented in Section~\ref{sec:applications}. 
Some longer proofs and auxiliary results are collected in the Appendix.

\section{Preliminaries}\label{sec:pre}

We fix an atomless probability space $(\Omega,{\cal F},P)$
and denote by $\E{\cdot}$ and $\var{\cdot}$ the expectation operator and the variance, respectively, with respect to the reference probability measure $P$. Likewise, we write $\EQ{\cdot}$ for the expectation operator with respect to a probability measure $Q$.
%For any random variable $X$ defined on this probability space, let
%\begin{equation}
%q_X(t):=\inf\{m\in \R | P[X\leq m]\geq t\}, \quad t\in [0,1],
%\label{eq:VaR}
%\end{equation}
%be its (left-continuous) quantile function,
%where $\inf \emptyset:=\infty$.

We consider preferences on payoffs that can be numerically represented by $\Ucal\circ u$ where $u$ is a utility function on the real numbers and the coherent criterion %$\Ucal$ is essentially a robust expectation. 
%More precisely we assume throughout this paper that 
$\Ucal:L^1:=L^1(\Omega,\mathcal{F},P)\to \R$ satisfies the following conditions:
\begin{itemize}
     \item[i)] superadditivity: $\Ucal(X+Y)\geq \Ucal(X)+\Ucal(Y)$ for all $X,Y\in L^1$,
    \item[ii)] positive homogeneity: $\Ucal(tX)=t\,\Ucal(X)$ for all $t\geq 0$ and $X\in L^1$,
    \item[iii)] monotonicity: $\Ucal(X)\leq \Ucal(Y)$ whenever $X,Y\in L^1$ are such that $X\leq Y$ ($:\Leftrightarrow P(X\leq Y)=1$),
    \item[iv)] cash-additivity: $\Ucal(X+a)=\Ucal(X)+a$ for all $X\in L^1$, $a\in \R$,
    \item[v)] law-invariance: $\Ucal(X)=\Ucal(Y)$ whenever $X\in L^1$ and $Y\in L^1$ share the same distribution under $P$ denoted by  $X\stackrel{d}{=}Y$, 
    \item[vi)] continuity: $\Ucal(X_n)\to \Ucal(X)$ whenever the sequence $(X_n)_{n\in \N}\subset L^1$ and $X\in L^1$ satisfy  $\lim_{n\to \infty}\E{|X_n-X|}=0$.
\end{itemize}

Note that $\Ucal$ is normalized ($\Ucal(0)=0$) and satisfies $\Ucal(a)=a$ for any constant $a\in\R$. %Examples of $\Ucal$ are  any law-invariant robust expectation 
An application of the Fenchel-Moreau theorem shows that $\Ucal$ satisfies conditions i)--vi) if and only if $\Ucal$ is a robust expectation \begin{equation}\label{eq:robust:exp}
\Ucal(X)={\rm E}_\mathcal{M}[X]:=\inf_{Q\in \mathcal{M}}\EQ{X},\end{equation} where $\mathcal{M}$ is a non-empty law-invariant set of probability measures on $(\Omega, \mathcal{F})$ that are absolutely continuous and have uniformly bounded densities with respect to $P$, see e.g.,\ \cite[Corollary~4.2]{CL2009}. 
Law-invariance of $\mathcal{M}$ means that for all $Q,Q'\ll P$ with $dQ/dP\stackrel{d}{=}dQ'/dP$ we have $Q\in \mathcal{M}$ if and only if $Q'\in \mathcal{M}$, see  \cite[Corollary 5.2]{BKMS21}. 
%In this case, $\Ucal={\rm E}_{\mathcal{M}}$ also satisfies 
%\begin{itemize}
%    \item[vi)] monotonicity: $\Ucal(X)\leq \Ucal(Y)$ whenever $X,Y\in L^1$ are such that $X\leq Y$ ($:\Leftrightarrow P(X\leq Y)=1$).
%\end{itemize}
%Our main result, Theorem~\ref{thm:main:1}, will not require monotonicity. However, most of the discussion  will be for monotone $\Ucal$. %An application of the Fenchel-Moreau theorem shows that if $\Ucal$ satisfying condtition i)--v) is also monotone, then $\Ucal$ is indeed a robust expectation of type \eqref{eq:robust:exp}, see e.g.\ \cite[Corollary~4.2]{CL2009}. 

As regards the utility function $u$, we assume that $u:\R\to \R\cup\{-\infty\}$ is a concave utility function that is increasing and strictly increasing on its domain $\operatorname{dom}u:=\{x\in \R\mid u(x)>-\infty\}$. 
Note that $u$ is continuous on the interior of $\operatorname{dom}u$, see \cite[Proposition A.4]{FS11}.
Allowing $u$ to take the value $-\infty$ enables us to incorporate utility functions such as the power utility or logarithmic utility, which are simply set to equal $-\infty$ outside their domain $[0,\infty)$ or $(0,\infty)$, respectively. 

Note that by Jensen's inequality $\EQ{u(X)}\leq u(\EQ{X})$ for any $X\in L^1$ also when $u(X)$ takes the value $-\infty$ with positive probability under $Q$. 
In particular, $\EQ{u(X)}\in \R\cup\{-\infty\}$ is well-defined for any $X\in L^1$ and thus $\Ucal(u(X))= \inf_{\Q\in \Mcal}\EQ{u(X)}$ is also well-defined, possibly taking the value $-\infty$. 
We will thus view $\Ucal\circ u$ as a function mapping $L^1$ to $\R\cup\{-\infty\}$. %Note that, due to concavity of $u$, $$\Ucal\circ u(X)= \inf_{Q\in \mathcal{M}}\EQ{u(X)}$$ is well-defined for every $X\in L^1$, possibly taking the value $-\infty$.
As we do not require strict concavity, the linear case $u(x)=\operatorname{id}(x):=x$, $x\in \R$, in case of which $\Ucal\circ u = \Ucal$, is also part of our analysis. 
However, in general, if $u\not \equiv \operatorname{id}$, $\Ucal\circ u$ will not equal $\Ucal$ and will thus in particular not be superadditive, positively homogeneous, or cash-additive. 
Clearly, $\Ucal\circ u$ preserves law-invariance, and a type of continuity, namely $\Ucal(u(X_n))\to \Ucal(u(X))$ whenever the sequence $(X_n)_{n\in \N}\subset L^1$ and $X\in L^1$ satisfy $\lim_{n\to \infty}\E{|X_n-X|}=0$ and $(u(X_n))_{n\in \N}\subset L^1$ is uniformly integrable. 
Moreover, superadditivity of $\Ucal$ implies concavity of $\Ucal\circ u$.   

Preferences represented by $\Ucal\circ u$ are sometimes referred to as probabilistically sophisticated Gilboa-Schmeidler (\cite{GS1989}) preferences, and cover the popular rank-dependent utility (RDU) model (Quiggin \cite{Q82}), %namely when $\Mcal=\{\mu\}$.
which in turn encompasses Yaari's \cite{Y87} dual theory of choice under risk and the expected utility model as special cases.
For the connection between law-invariance and probabilistic sophistication introduced by Machina and Schmeidler \cite{MS92}, we refer to Marinacci \cite{M02}, Maccheroni, Marinacci and Rustichini \cite{MMR06}, Strzalecki \cite{S11}, and Ravanelli and Svindland \cite{RS14}.

\setcounter{equation}{0}

\section{Limiting Behavior of the Risk Premium under Linear Risk Sharing}\label{sec:main}

\subsection{Risk Premium}
Let  $v\geq 0$ denote an economic agent's initial wealth level. Consider a risk $X\in L^1$. %such that $\operatorname{Im}(X):=\{X(\omega)\mid \omega\in \Omega\}\subset \operatorname{dom}u$ and $u(X)\in L^1$, and let  $v\geq 0$ denote an economic agent's initial wealth level.
Then the risk premium $\pi(v,X)$ (if it  exists) with respect to $\Ucal\circ u$ is given by any real number $\pi(v,X)$ such that $v+ \E{X}-\pi(v,X)\in \dom u$ and
\begin{equation}
\Ucal\left(u\left(v+X\right)\right)=u\left(v+ \E{X}-\pi(v,X)\right).
\label{eq:equivutility}
\end{equation}
The risk premium is such
that the economic agent is indifferent between bearing the risk $X$ or receiving the expectation $\E{X}$ of the risk and paying the risk premium.  
As the utility function $u$ is assumed to be strictly increasing on its domain and thus invertible, if the risk premium exists, then 
\begin{equation}
\pi(v,X):=v+\E{X}-u^{-1}\left(\Ucal\left(u\left(v+X\right)\right)\right),
\label{eq:riskpremium}
\end{equation}  
where $u^{-1}:\operatorname{Im}(u):=\{u(x)\mid x\in \operatorname{dom}u \}\to \R$ denotes the inverse function of $u$, that is, $u^{-1}\circ u(x)=x$ for all $x\in \operatorname{dom}u$.
Henceforth, we let $v= 0$ without losing generality, and set $\pi(X):=\pi(0,X)$. Indeed, note that $\pi(v+X)=\pi(v,X)$, and therefore the limiting behavior of the risk premium  for non-trivial $v$ follows from the limiting behavior of the risk premium with zero initial capital by adding $v$ to the risk.  
We will call a risk $X\in L^1$ {\it compatible} with $(\Ucal, u)$ if the risk premium $\pi(X)$ exists. 

\subsection{Linear Risk Sharing}
We consider a large portfolio consisting of $n$ risks modeled as independent random variables $X_1,\ldots,X_n$, $n\in\N$, on $(\Omega,\F,P)$ and denote by $S_n:=X_1+\ldots+X_n$ the aggregate position. 
We do not limit the discussion to the classical i.i.d.\ setting, but also  allow for heterogeneous, however independent, risks. 
The classical framework with identically distributed risks appears as a special case.

A risk sharing rule $(Y_1,\ldots, Y_n)$ is a decomposition $S_n=Y_1+\ldots+Y_n$ such that the random variable $Y_i$ is reallocated to the risk $i$. In practice, linear risk sharing rules of type 
\begin{equation}
\label{eq:linearrisksharingrule}
Y^n_i:=\E{X_i}+a_n^i(S_n-\E{S_n}),
\end{equation}
given by numbers $a_n^i\geq 0$, $i=1,\ldots,n$, $n\in \N$, such that $\sum_{i=1}^na_n^i=1$, are of particular importance since they can readily be  implemented, and at the same time are easy to communicate to different stakeholders. Here the $a_n^i$ may depend on the risks $X_i$ and $S_n$, $i=1,\ldots,n$.
Prominent examples of linear risk sharing rules include (e.g., \cite{DR21}; assuming that all quantities in the following are well-defined):
\begin{itemize}
    \item \emph{Proportional rule:}
    $$a^{i,\text{prop}}_n:=\frac{\E{X_i}}{\E{S_n}}. $$
    \item \emph{Linear regression rule:}
    $$ a^{i,\text{reg}}_n:=\frac{\var{X_i}}{\sigma^2_n}\quad\mbox{where $\sigma^2_n:=\sum_{i=1}^n \operatorname{Var}(X_i)$}.$$
    \item \emph{Mean-variance rule:}
    $$a^{i,\text{mv}}_n:=\beta a^{i,\text{prop}}_n+(1-\beta)a^{i,\text{reg}}_n,\quad\beta\in(0,1).$$
\end{itemize}

\begin{remark}[Homogeneous sequence of random variables]\label{rem:homogeneouscase}
Suppose that the independent random variables $(X_i)_{i\in\mathbb{N}}\subset L^2$ are homogeneous in the sense that, for all $i\in\N$, $\E{X_i}\equiv \E{X_1}$ and $\var{X_i}\equiv\var{X_1}$. 
Then the proportional, the linear regression, and the mean-variance risk sharing rules coincide and correspond to the equal risk sharing rule, i.e.,
$$a^{i,\rm{prop}}_n=a^{i,\rm{reg}}_n=a^{i,\rm{mv}}_n=\tfrac{1}{n}\quad \mbox{and}\quad Y^n_i=\frac1nS_n\quad  \mbox{for all   $i=1,\ldots,n$}. $$
\end{remark}

\subsection{Asymptotic Behavior}
In the following, for a given linear risk sharing rule (\ref{eq:linearrisksharingrule}), we analyze the decay of the risk premium
$$\pi(Y^n_i)=\pi(\E{X_i}+a_n^i(S_n-\E{S_n}))$$
for all $i\leq n$ if the portfolio size $n$ tends to $\infty$. 

\begin{theorem}\label{thm:main:1} Assume that $\mathcal{U}:L^1\to \R$ satisfies properties i)--vi). 
Assume further that $u$ and $u^{-1}$ are continuously differentiable on the interior of their respective domains and that the sequence $(X_i)_{i\in \N}\subset L^2$ obeys
\begin{itemize}
\item  $(X_i)_{i\in \N}$ are independent,
%\item common mean $\E{X_i}=\E{X_1}=:\mu$  and variance $\operatorname{Var}(X_i)=\operatorname{Var}(X_1)=:\sigma^2$ for all $i\in \N$,
\item the strong law of large numbers:  $\frac1n(S_n-\E{S_n})\to 0$ $P$-a.s., 
\item the central limit theorem: $ S_{n}^{*}:=\frac{1}{\sigma_n} (S_n-\E{S_n}) \stackrel{d}{\to}N(0,1)$ where $\sigma^2_n=\sum_{i=1}^n \operatorname{Var}(X_i)$,
\item $\operatorname{Im}(X_i)\subset \operatorname{int}\operatorname{dom} u$ for all $i\in \N$.
\end{itemize} 
Consider a linear risk sharing rule of type (\ref{eq:linearrisksharingrule}). Further assume that for $i\in \N$ 
\begin{itemize}
\item $Y^n_i$ is compatible with $(\Ucal,u)$ and  $\operatorname{Im}(Y^n_i)\subset \operatorname{int}\operatorname{dom} u$ for all $n\in \N$ such that $n\geq i$, 
\item there is $G_i\in L^2$ such that
$| u'(Y^n_i)|\leq G_i$
for all $n\in \N$ such that $n\geq i$, 
\item  and that
\begin{equation}
\label{eq:condconvergence1} \sup_{n\in \N}a_n^in<\infty \quad \mbox{and} \quad 
c_i:= \limsup_{i\leq n\to \infty }\sqrt{n} a_n^i \sigma_n<\infty.
\end{equation}
\end{itemize}
Then we have that  
% 
 %$u(\E{X_i}+a_n^i(S_n-\E{S_n}))\in L^1$ for all $i,n\in \N$ and that there is $G_i\in L^2$ such that $|u'(\min\{\E{X_i},\E{X_i}+a_n^i(S_n-\E{S_n})\})|\leq G_i$ for all $n\in \N$ and $i\in \N$.  
 %Then, 
\begin{align}\label{eq:conv:rate}
\limsup_{i\leq n\rightarrow\infty}\sqrt{n}\;\pi(Y^n_i) 
\leq -c_i \,\Ucal \left(Z\right),
\end{align}
where $Z$ is a standard normal random variable under $P$. If \begin{equation}\label{eq:ci:cond}\limsup_{i\leq n\to \infty }\sqrt{n} a_n^i \sigma_n=\liminf_{i\leq n\to \infty }\sqrt{n} a_n^i \sigma_n,\end{equation} then \begin{align}\label{eq:conv:rate:lim}
\lim_{i\leq n\rightarrow\infty}\sqrt{n}\;\pi(Y^n_i) 
=-c_i \,\Ucal \left(Z\right).
\end{align}
Moreover, if $\Ucal(X)=\E{X}$, $u$ is two times continuously differentiable on the interior of its domain, and the $X_i$, $i\in \N$, are uniformly bounded, then  %i.e., for 
%$$\pi(\E{X_i}+a^{i}_n(S_n-\E{S_n}))=\E{X_i}-u^{-1}\left(\E{u(\E{X_i}+a^{i,\cdot}_n(S_n-\E{S_n}))}\right)$$
%it holds that
\begin{equation}  \label{eq:conv:rate:no:amb}
 %\limsup_{i\leq n\uparrow\infty}\sqrt{n}\pi(\E{X_i}+a^{i,\cdot}_n(S_n-\E{S_n}))&\leq& c_i,\\
%\lim_{i\leq n\to \infty}n\pi(\E{X_i}+a^{i,\cdot}_n(S_n-\E{S_n}))
\limsup_{i\leq n\rightarrow\infty}n\;\pi(Y^n_i)  \leq \frac{1}{2}R(\E{X_i})(c_i)^2,
\end{equation}
and in case of \eqref{eq:ci:cond}  $$\lim_{i\leq n\rightarrow\infty}n\;\pi(Y^n_i)  =\frac{1}{2}R(\E{X_i})(c_i)^2,$$
where $R(x)=-u''(x)/u'(x)$ denotes the Arrow-Pratt coefficient of absolute risk aversion. 
\end{theorem}
The proof of Theorem~\ref{thm:main:1} is provided in Appendix~\ref{sec:proof:mainthm}.
Note that in the homogeneous case of Remark~\ref{rem:homogeneouscase} and for $a^{i}_n=1/n$ we have that $\sqrt{n}a^{i}_n\sigma_n=\sqrt{\var{X_1}}$ for all $i,n\in\N$, and thus $c_i=\sqrt{\var{X_1}}$ for all $i\in\N$. 
Then, Theorem~\ref{thm:main:1} simplifies to:

\begin{corollary}\label{cor:main:iid} 
Under the conditions stated in Theorem~\ref{thm:main:1}, suppose further that the $(X_i)_{i\in\mathbb{N}}$ are homogeneous in the sense that, for $i\in\N$, $\E{X_i}\equiv \E{X_1}$ and $\var{X_i}\equiv\var{X_1}$. Then
\begin{align}\label{eq:conv:rate:iid}
\lim_{n\rightarrow\infty}\sqrt{n}\;\pi(S_n/n) 
=-\sqrt{\var{X_1}} \,\Ucal \left(Z\right).
\end{align}
Moreover, if $\Ucal(X)=\E{X}$ and $u$ is two times continuously differentiable on the interior of its domain, then  %i.e., for 
%$$\pi(\E{X_i}+a^{i}_n(S_n-\E{S_n}))=\E{X_i}-u^{-1}\left(\E{u(\E{X_i}+a^{i,\cdot}_n(S_n-\E{S_n}))}\right)$$
%it holds that
\begin{equation} \label{eq:conv:rate:no:amb:iid}
 %\limsup_{i\leq n\uparrow\infty}\sqrt{n}\pi(\E{X_i}+a^{i,\cdot}_n(S_n-\E{S_n}))&\leq& c_i,\\
\lim_{i\leq n\to \infty}n\pi(S_n/n)=\frac{1}{2}R(\E{X_1})\var{X_1}.
\end{equation}
\end{corollary}

\begin{remark}
An independent sequence $(X_i)_{i\in \N}\subset L^2$ meets the conditions required in  Theorem~\ref{thm:main:1} whenever this sequence satisfies both
\begin{itemize}
\item Kolmogorov's condition for the strong law of large numbers: \begin{equation*}
        \sum_{i\in \N} \frac{\operatorname{Var}(X_i)}{i^2}<\infty, \quad \mbox{see \cite[VII.8 Theorem~3]{Feller1970}},\end{equation*}
    \item Lindeberg's condition for the central limit theorem: for all $t>0$ \begin{equation*}\lim_{n\to \infty}\frac{1}{\sigma_n^{2}}\sum_{i=1}^n \E{|X_i-\E{X_i}|^{2}1_{\{ |X_i-\E{X_i}|\geq t\sigma_n\}}}=0, \quad \mbox{see \cite[VIII.4 Theorem~3]{Feller1970}}.
    \end{equation*}
    %Lyapunov's condition for the central limit theorem: there is $\delta>0$ such that \begin{equation*}\lim_{n\to \infty}\frac{1}{\sigma_n^{2+\delta}}\sum_{i=1}^n \E{|X_i-\E{X_i}|^{2+\delta}}=0, \quad \mbox{see ??}.\end{equation*}
\end{itemize}
\end{remark}

%\begin{remark}\label{rem:2ndorder} {\color{red} Not nice to have the following in a remarks, since it is part of the main stroy. However, we need to refer to this later. }
%Proposition 2.4 and Theorem 2.5 in \cite{KLS16} describe the decay of the risk premium for i.i.d.\ random variables with respect to expected utility, i.e., $\Ucal(X)=\E{X}$. 
%Under the assumption $\lim_{i\leq n\to \infty}\sqrt{n}a^{i}_n\sigma_n=c_i$, these results can be readily generalized to the case of heterogeneous losses, i.e., for 
%$$\pi(\E{X_i}+a^{i}_n(S_n-\E{S_n}))=\E{X_i}-u^{-1}\left(\E{u(\E{X_i}+a^{i,\cdot}_n(S_n-\E{S_n}))}\right)$$
%it holds that
%\begin{equation} \label{eq:exp:utility:rate}
 %\limsup_{i\leq n\uparrow\infty}\sqrt{n}\pi(\E{X_i}+a^{i,\cdot}_n(S_n-\E{S_n}))&\leq& c_i,\\
%\lim_{i\leq n\to \infty}n\pi(\E{X_i}+a^{i,\cdot}_n(S_n-\E{S_n}))=\frac{1}{2}R(\E{X_i})(c_i)^2,
%\end{equation}
%where $R(x)=-u''(x)/u'(x)$ denotes the Arrow-Pratt coefficient of absolute risk aversion. 
The rate $n$ in \eqref{eq:conv:rate:no:amb} shows that if $\Ucal(\cdot)=\E{\cdot}$, then the limit in \eqref{eq:conv:rate} should equal $0$, which is indeed the case since $\Ucal(Z)=\E{Z}=0$. 
In the following, we will see that the rate of convergence $n$ as in \eqref{eq:conv:rate:no:amb} is essentially obtained if and only if $\Ucal(\cdot)=\E{\cdot}$.   
%To this end, suppose the criterion $\Ucal$ is also monotone, see property vi). 
%Then, as already mentioned, by the Fenchel-Moreau theorem $\Ucal$ admits a dual representation as a robust expectation $\Ucal={\rm E}_\mathcal{M}$ as in \eqref{eq:robust:exp}, where $\mathcal{M}\subset \{Q \mid Q\ll P\,\mbox{and}\,  dQ/dP\in L^\infty\}$ is law-invariant,  see, e.g., \cite[Corollary~4.2]{CL2009}. 
%\begin{equation}\label{eq:dual:rep}
%    \Ucal(X)=\inf_{Q\in \mathcal{M}} \EQ{X}, \quad X\in L^1,
%\end{equation}
%where $$\mathcal{M}:=\{Q\mid Q\ll P\; \mbox{s.t.}\; dQ/dP\in L^\infty\; \mbox{and} \;\forall Y\in L^1: \EQ{Y}\geq \Ucal(Y)\}$$ is a law-invariant set of probability measures, see e.g.\ \cite[Proposition 4.15]{FS11} and \cite[Corollary 5.2]{BKMS21}. 
To this end, for a random variable $X$, we denote by
\begin{equation}
q_X(t):=\inf\{m\in \R | P[X\leq m]\geq t\}, \quad t\in [0,1],
\label{eq:VaR}
\end{equation}
its (left-continuous) quantile function, where $\inf \emptyset:=\infty$. Let \begin{equation}\label{eq:AM:def}\mathcal{A}(\mathcal{M}):=-\Ucal\left(Z\right) =\sup_{Q\in \mathcal{M}} \EQ{-Z}= \sup_{Q\in \mathcal{M}}\int_0^1q_Z(s)q_{\frac{dQ}{dP}}(s) \, ds,\end{equation} where $Z$ is a standard normal random variable and the set of probabilities $\Mcal$ is given by a dual representation \eqref{eq:robust:exp}.
Note that the last equality in \eqref{eq:AM:def} follows from law-invariance of $\mathcal{M}$ and \cite[Lemma 4.60]{FS11}, realling that $q_{-Z}=q_{Z}$ due to the symmetry of the normal distribution. 
Even though the $Q$ probabilities in $\mathcal{M}$ have bounded densities with respect to $P$, in the following we extend the discussion by relaxing this boundedness condition and only asking for \begin{equation}\label{eq:quantile:cond}\int_0^1 q_Z(s)q_{\frac{dQ}{dP}}(s)\, ds<\infty.\end{equation} Denote by $\mathfrak{P}$ the set of all probability measures $Q$ on $(\Omega,\mathcal{F})$ such that $Q\ll P$ and \eqref{eq:quantile:cond} is satisfied. For any non-empty $\mathcal{M}\subset \mathfrak{P}$ %which is non-empty and law-invariant 
let $$\mathcal{A}(\mathcal{M}):= \sup_{Q\in \mathcal{M}} \int_0^1q_Z(s)q_{\frac{dQ}{dP}}(s) \, ds.$$

We state the following proposition:
\begin{proposition}\label{prop:risk:aversion} Let $\mathcal{M},\widetilde{\mathcal{M}}\subset \mathfrak{P}$ be non-empty and law-invariant. 
Then: 
\begin{enumerate}
\item $\mathcal{A}(\mathcal{M})\geq 0$.
\item $\mathcal{M}\subseteq \widetilde{ \mathcal{M}}$ implies $\mathcal{A}(\mathcal{M})\leq \mathcal{A}(\widetilde{\mathcal{M}})$.
\item $\mathcal{A}(\mathcal{M})=0\; \Leftrightarrow \;\mathcal{M}=\{P\}\; \Leftrightarrow\; \rm{E}_\mathcal{M}[\cdot]= \E{\cdot}$.
\end{enumerate}
\end{proposition}

\begin{proof} 2. is obvious. 
Next, we prove 1.: Let $Q\in \mathfrak{P}$. 
Using the symmetry of the quantile function of the standard normal distribution, we find that $$\int_0^1q_Z(s)q_{\frac{dQ}{dP}}(s)\,ds=\int_0^{\frac12}q_Z( \frac12 + s) \left(q_{\frac{dQ}{dP}}(\frac12 +s)-q_{\frac{dQ}{dP}}(\frac12 -s)\right)\, ds,$$ where the function \begin{equation}\label{eq:help:AM}\left(0,\frac12 \right)\ni s\mapsto  q_{\frac{dQ}{dP}}(\frac12 +s)-q_{\frac{dQ}{dP}}(\frac12 -s)\end{equation} is non-negative and $q_Z(\frac12 +s)$ is strictly positive for $s\in (0,\frac12)$. 
Hence, it follows that \begin{equation}\label{eq:help:AM2}\int_0^1q_Z(s)q_{\frac{dQ}{dP}}(s)\,ds\geq 0\end{equation} for all $Q\in \mathfrak{P}$, and therefore $\mathcal{A}(\mathcal{M})\geq 0$. 
As for 3., suppose that $\mathcal{A}(\mathcal{M})=0$. 
By \eqref{eq:help:AM2} this implies that indeed $\int_0^1q_Z(s)q_{\frac{dQ}{dP}}(s)\,ds= 0$ for all $Q\in \mathcal{M}$. 
But that can only happen if the function given in \eqref{eq:help:AM} equals the constant zero function. 
The latter however implies that $Q=P$, so $\mathcal{M}=\{P\}$.
The other implications are easily verified. 
\end{proof}

Proposition~\ref{prop:risk:aversion} implies that a convergence rate $n$ as in \eqref{eq:conv:rate:no:amb} of  Theorem~\ref{thm:main:1} essentially necessitates $\Ucal(\cdot)=\E{\cdot}$. 
More precisely, assume that  $c_i>0$. If $\lim_{i\leq n\rightarrow\infty}n\;\pi(Y^n_i)$ exists as a real number, we must have $\lim_{i\leq \sqrt{n}\rightarrow\infty}\sqrt{n}\;\pi(Y^n_i)=0$, and hence $\Ucal(Z)=0$, that is, $\mathcal{A}(\Qcal)=0$, so $\Ucal=\E{\cdot}$.    

\begin{remark}
\label{thm:first-second-order-risk-aversion} The dichotomy of Theorem~\ref{thm:main:1}, Eqns.~\eqref{eq:conv:rate} and \eqref{eq:conv:rate:no:amb}, is connected to, and suitably modifies, first- and second-order risk aversion introduced by Segal and Spivak \cite{SS90}; see also Lang \cite{L17} and Eeckhoudt and Laeven \cite{EL21}. 
Segal and Spivak \cite{SS90} show that the risk premia have distinct limiting behavior for `small' risks under RDU and under expected utility: under the RDU model, risk aversion is a first-order phenomenon, whereas under the expected utility model risk aversion is a second-order phenomenon; see also Eeckhoudt and Laeven \cite{EL21}, Section~5.3.
%
%In this section, we analyze the limiting measure $-\mathcal{U}(Z)$ given in \eqref{eq:conv:rate} of Theorem~\ref{thm:main:1}. 
%We also relate the dichotomy, observed upon comparing Theorem~\ref{thm:main:1} and Remark~\ref{rem:2ndorder}, to first- and second-order risk aversion introduced by Segal and Spivak \cite{SS90}; see also Lang \cite{L17} and Eeckhoudt and Laeven \cite{EL21}.  
%
%Indeed, the insightful analysis of Segal and Spivak \cite{SS90} shows that the risk premia have distinct limiting behavior for `small' risks under RDU and under expected utility:
% 
%Whether risk aversion is of order one or of order two also is known to have important economic and financial implications, e.g., for optimal portfolio choice, see ??.
%
Small risks are risks of type $v+tY$, where $v\geq 0$ denotes an initial capital and $Y$ is sufficiently integrable, for instance normally distributed with mean $\E{Y}=0$ and variance $\operatorname{Var}(Y)=1$. 
In fact, similar analysis as in the proof of Theorem~\ref{thm:main:1} shows that 
\begin{equation}\label{eq:first:order:help}
\lim_{t\downarrow 0}\frac1t u^{-1}\left(\Ucal(u(v+tY))\right)=  -\Ucal(Y).\end{equation} 
This corresponds to first-order behavior in the sense of \cite{SS90} unless $\Ucal(Y)=0$. 
Indeed, if the rate $t$ is identified with $\frac{1}{\sqrt{n}}$, then \eqref{eq:first:order:help} equals $$\lim_{n\to \infty}\sqrt{n} u^{-1}\left(\Ucal(u(v+\frac{Y}{\sqrt{n}}))\right)= \lim_{n\to  \infty}\sqrt{n} \pi\left(v,\frac{Y}{\sqrt{n}}\right)=- \Ucal(Y).$$
Let $(X_i)_{i\in \N}$ be i.i.d.\,with zero mean and $\operatorname{Var}(X_1)>0$. 
For large $n\in \N$ we have that $\frac{S_n}{\sqrt{n}}$ is almost normal with zero mean and variance $\operatorname{Var}(X_1)$. 
Hence,  by law-invariance, continuity, and positive homogeneity of $\Ucal$, 
$$\lim_{n\to \infty}\sqrt{n} \pi\left(v,\frac{S_n}{n}\right) = \lim_{n\to \infty}\sqrt{n} \pi\left(v,\frac{\sqrt{\operatorname{Var}(X_1)} Z}{\sqrt{n}}\right)= -\sqrt{\operatorname{Var}(X_1)} \Ucal(Z),$$ where the random variable $Z$ is standard normal under $P$. 
Therefore, \eqref{eq:conv:rate:iid}, and also more generally \eqref{eq:conv:rate}, may be seen as corresponding to first-order risk aversion provided $\Ucal(Z)\neq 0$. 
If $\Ucal(Z)=0$, that is, $\Ucal(\cdot)=\E{\cdot}$ according to Proposition~\ref{prop:risk:aversion}, we have \eqref{eq:conv:rate:no:amb}, which is---similar to above---naturally identified with second-order risk aversion.
%
%This leads to the following adaptation of first- and second-order risk aversion to our framework:
%\begin{definition} 
%We say that the preferences given by $\Ucal(u(\cdot))$ display first-order risk aversion if $\lim_{n\rightarrow\infty}\sqrt{n}\,\pi(S_n/n)\neq 0 $ (and in particular the limit exists) for all i.i.d.\ sequences $(X_i)_{i\in \N}\subset L^\infty$ such that $S_n/n$ is compatible with $(\Ucal,u)$ for all $n$. 
%The preferences exhibit second-order risk aversion if $\lim_{n\rightarrow\infty} n\,\pi(S_n/n)\neq 0 $ for all i.i.d.\ sequences $(X_i)_{i\in \N}\subset L^\infty$ such that $S_n/n$ is compatible with $(\Ucal,u)$ for all $n$ (and thus $\lim_{n\rightarrow\infty}\sqrt{n}\;\pi(S_n/n)=0 $).   
%\end{definition}
%
%
%
%As a consequence, any agent with preferences given by a law-invariant robust expected utility ${\rm E}_\mathcal{M}[u(\cdot)]$  with $\Mcal\neq \{P\}$ exhibits first-order risk aversion, independent of the utility function $u$ as long as $u$ satisfies the (smoothness) conditions of Theorem~\ref{thm:main:1}. 
%Any expected utility agent $\E{u(\cdot)}$, however, displays second-order risk aversion. 
%Note that Segal and Spivak \cite{SS90} also find first-order risk aversion in case of expected utility, but this requires non-smooth utility functions $u$, which are not discussed in this paper. 
\end{remark}

\begin{remark}
By analogy to $R(\E{X})$, which is a measure of second-order risk aversion, $\mathcal{A}(\mathcal{M})$ may be interpreted as a measure of first-order risk aversion.    
The measure $\mathcal{A}(\mathcal{M})$ may be evaluated analytically in a wide variety of cases, e.g., under RDU with probability weighting/distortion function of the power type (\cite{CM09,ELS20}) or induced by the Esscher-Girsanov transform (\cite{GL08,LO10}).
\end{remark}

We end this section by showing that it is possible that the right-hand side in \eqref{eq:conv:rate} equals zero, even though $\Ucal(Z)\neq 0$. 
That, of course, requires $c_i=0$.

\begin{example} Let $(X_i)_{i\in \N}$ be independent, satisfying the central limit theorem (Lindeberg condition), such that 
 $\E{X_i}=\ln(i)$ and $\var{X_i}=\ln(\sqrt{i})$, $i\in\N$. 
In this case, we have $\E{S_n}=\ln(n!)$ and $\sigma_n^2=\var{S_n}=\tfrac{1}{2}\ln(n!)$. 
But this yields
$$ \sqrt{n}a^{\rm{prop},i}_n\sigma_n=\sqrt{n}\frac{\ln(i)}{\ln(n!)}\sqrt{\tfrac{1}{2}\ln(n!)}=\sqrt{\frac{n}{\ln(n!)}}\sqrt{\tfrac{1}{2}}\ln(i),$$
i.\,e., $c_i^{\rm{prop}}=\lim_{n\uparrow\infty}\sqrt{n}a^{\rm{prop},i}_n\sigma_n=0$. Similarly, we obtain
$$ \sqrt{n}a^{\rm{reg},i}_n\sigma_n=\sqrt{n}\frac{\ln(\sqrt{i})}{\tfrac{1}{2}\ln(n!)}\sqrt{\tfrac{1}{2}\ln(n!)}=\sqrt{\frac{n}{\ln(n!)}}\sqrt{2}\ln(\sqrt{i}),$$
i.\,e., $c_i^{\rm{reg}}=\lim_{n\uparrow\infty}\sqrt{n}a^{\rm{reg},i}_n\sigma_n=0$.
Note that the strong law of large numbers holds due to Kolmogorov\rq s condition
$$\sum_{i\in\mathbb{N}}\frac{\var{X_i}}{i^2}=\frac{1}{2}\sum_{i\in\mathbb{N}}\frac{\ln(i)}{i^2}\leq \frac{1}{2}\sum_{i\in\mathbb{N}}\frac{\sqrt{i}}{i^2}=\frac{1}{2}\sum_{i\in\mathbb{N}}\frac{1}{i^{3/2}}<\infty,$$
in view of the convergence of the generalized harmonic series.
\end{example}

\setcounter{equation}{0}

\section{Pareto Optimality of Linear Risk Sharing}\label{sec:Pareto}

Since we assume linear risk sharing in Section~\ref{sec:main}, a natural question is whether such risk sharing rules are Pareto optimal when all agents share the same preferences given by $\Ucal\circ u$. 
In this section, we address this question.  
%To this end, we assume that $\Ucal$ is monotone, so that $\Ucal=\mathrm{E}_\Qcal $ as in \eqref{eq:robust:exp}.  
Without loss of generality, in the following, we will always work with the maximal set $$\Qcal(\Ucal)=\{Q\ll P\mid  dQ/dP\in L^\infty, \forall Y\in L^1: \EQ{Y}\geq \Ucal(Y)\},$$ for which we obtain $\Ucal=\mathrm{E}_{\Qcal(\Ucal)}$. 

%Also recall that $\Ucal\circ u$ is concave.
%By upper semi-continuity of $u$ it follows that indeed $\Ucal(u(X))\leq u(\Ucal(X))$
For given $n\in \N$ and $X\in L^1$, consider the set of allocations
\begin{equation*}
\mathbb{A}_n(X):=\left\{(Z_{1},Z_{2}, \ldots, Z_n)\in (L^1)^n \mid \sum_{i=1}^nZ_{i}=X\right\}.    
\end{equation*}
We recall the definition of Pareto optimality:

\begin{definition}
An allocation $(Z_{1},Z_{2}, \ldots, Z_n)\in\mathbb{A}_n(X)$ such that $\mathcal{U}(u(Z_{i}))>-\infty$ for all $i=1,\ldots,n$ is called Pareto optimal if for all $(Y_{1},Y_{2}, \ldots, Y_n)\in\mathbb{A}_n(X)$ we have that $\mathcal{U}(u(Y_{i}))\geq \mathcal{U}(u(Z_{i}))$ for all $i=1,\ldots, n$ implies that
$\mathcal{U}(u(Y_{i}))=\mathcal{U}(u(Z_{i}))$ for all $i=1,\ldots, n$. 
\end{definition}

The following lemma, which states that the equal risk sharing rule is Pareto optimal, is folklore, but we provide a short proof of the sake of completeness. 

%$\operatorname{Im}(X)\subset \operatorname{int}\operatorname{dom} u$,
\begin{lemma}
For any $X\in L^1$, the equal risk sharing rule $(X/n,\ldots,X/n)$, allocating $X$ uniformly among the $n$ agents, is Pareto optimal.
\end{lemma}

\begin{proof}
Consider any $(Y_{1},\ldots,Y_{n})\in \mathbb{A}_n(X)$, s.t.
\begin{equation*}
\forall i\in \{1,\ldots,n \}\quad \Ucal(u(Y_{i}))\geq \Ucal(u(X/n)).
\end{equation*}
By concavity of $\Ucal\circ u$, 
\begin{equation*}
\frac{1}{n}\sum_{i=1}^{n}\Ucal(u(Y_{i}))\leq \Ucal\left(u(\frac{1}{n}\sum_{i=1}^{n}Y_{i})\right)=\Ucal(u(X/n)).
\end{equation*}
Hence, it follows that 
\begin{equation*}
\Ucal(u(Y_{i}))=\Ucal(u(X/n)),
\label{eq:equality}
\end{equation*}
for all $i=1,\ldots,n$.
Thus, $X/n$ is a Pareto optimal allocation of $X$.
\end{proof}

Therefore, if the sequence $X_i$, $i\in \N$, is homogeneous as in Remark~\ref{rem:homogeneouscase}, then the proportional, the linear regression, and the mean-variance risk sharing rules, all coinciding with the equal risk sharing rule,  are Pareto optimal.

A more general result on the existence and characterization of Pareto optima is the following Proposition~\ref{prop:Pareto}, which is a generalization of Borch's theorem (\cite{B60}). 
To this end, we recall that an allocation $(Z_{1},Z_{2}, \ldots, Z_n)\in\mathbb{A}_n(X)$ is Pareto optimal only if there exist Negishi weights $\lambda_1\geq  0,\ldots, \lambda_n\geq  0$ such that $\sum_{i=1}^n\lambda_i>0$ and  
\begin{equation} \label{eq:Pareto:SWF} \forall (Y_1,\ldots, Y_n)\in \mathbb{A}_n(X):  \sum_{i=1}^n\lambda_i\Ucal(u(Z_i))\geq \sum_{i=1}^n\lambda_i\Ucal(u(Y_i)).
\end{equation} 
In fact, \eqref{eq:Pareto:SWF} is also sufficient if all Negishi weights are strictly positive, that is, any allocation $(Z_{1},Z_{2}, \ldots, Z_n)\in\mathbb{A}_n(X)$ satisfying \eqref{eq:Pareto:SWF} with $\lambda_i>0$, $i=1,\ldots,n$, is Pareto optimal. 
Moreover, in case the criterion $\Ucal\circ u$ is strictly monotone, an allocation $(Z_{1},Z_{2}, \ldots, Z_n)\in\mathbb{A}_n(X)$ is Pareto optimal if and only if \eqref{eq:Pareto:SWF} holds. %Indeed, consider for example the case of expected utility $\EQ{u(\cdot)}$: If w.l.o.g.\ $\lambda_1=0$ and $\lambda_2>0$, then altering the allocation $(Z_{1},Z_{2}, \ldots, Z_n)\in\mathbb{A}_n(X)$ to $(\tilde Z_1, \tilde Z_2,\ldots, \tilde Z_n):=(Z_{1}-k,Z_{2}+k, \ldots, Z_n)$ for $k>0$ yields \begin{eqnarray*}\sum_{i=1}^n\lambda_i\EQ{u(\tilde Z_i)}&=& \lambda_2 \EQ{u(Z_2+k)} +\sum_{i=3}^n\lambda_i\EQ{u( Z_i)} \\&>&\lambda_2 \EQ{u(Z_2)} +\sum_{i=3}^n\lambda_i\EQ{u( Z_i)} \\ &=& \sum_{i=1}^n\lambda_i\EQ{u(Z_i)}\end{eqnarray*} since the strict monotonicity of $u$ implies that $\EQ{u(Z_2+k)}>\EQ{u(Z_2)}$.
%Hence, Negishi weights taking the value zero cannot characterize any Pareto optimum $(Z_{1},Z_{2}, \ldots, Z_n)$ in this case.

\begin{proposition}\label{prop:Pareto} Suppose that $u$ is differentiable on $\operatorname{int}\operatorname{dom}u$.
Let $X\in L^1$ and suppose that the allocation $(Z_1,\ldots, Z_n)\in \mathbb{A}_n(X)$ satisfies Borch's condition: 
\begin{equation}\label{eq:Borch}
    \mbox{there exist}\; \lambda_1,\ldots, \lambda_n>0 \; \mbox{such that }\; \lambda_1 u'(Z_1)=\ldots = \lambda_n u'(Z_n).
\end{equation}
(Here we, of course, implicitly assume that $\operatorname{Im}(Z_i)\subset\operatorname{int}\operatorname{dom}u $ so that $u'(Z_i)$ is well-defined for all $i=1,\ldots, n$.) 
Moreover, suppose that $(Z_1,\ldots, Z_n)$ admit a joint supergradient, that is, there is a probability measure $Q\in \Qcal(\Ucal)$ such that  
\begin{equation}\label{eq:supergrad:cond} \Ucal(u(Z_i))=\EQ{u(Z_i)} \quad \mbox{for all}\;  i=1,\ldots,n.\end{equation} 
Then $(Z_1,\ldots, Z_n)$ is Pareto optimal.
\end{proposition}

The following proof is a generalization of Borch's proof. 

\begin{proof}
Consider any other allocation $(Y_1,\ldots, Y_n)\in \mathbb{A}_n(X)$.
Borch's condition (see Borch \cite{B60} or recall the Kuhn-Tucker theorem \cite[Corollary 28.3.1]{Rock70}) ensures that 
$$\sum_{i=1}^n\lambda_i u(Z_i)\geq \sum_{i=1}^n\lambda_i u(Y_i)$$ and thus
$$\sum_{i=1}^n\lambda_i\EQ{u(Z_i)}\geq \sum_{i=1}^n\lambda_i\EQ{u(Y_i)}.$$ 
Hence, recalling that $\Ucal(\cdot)=\inf_{R\in \Qcal(\Ucal)}\mathrm{E}_{R}[\cdot]$,
\begin{equation*}
\sum_{i=1}^n\lambda_i\Ucal(u(Z_i)) = \sum_{i=1}^n\lambda_i\EQ{u(Z_i)}  \geq  \sum_{i=1}^n\lambda_i\EQ{u(Y_i)} \geq  \sum_{i=1}^n\lambda_i\Ucal(u(Y_i)),
\end{equation*}
and Pareto optimality of $(Z_1,\ldots, Z_n)$ follows.
\end{proof}

\begin{remark}
A closer inspection of the proof shows that in fact a more general version of Proposition~\ref{prop:Pareto} for concave instead of coherent (positively homogeneous and superadditive) $\Ucal_i$ (allowed to depend on $i$) is true, which may be of independent interest and is provided in Appendix~\ref{app:Pareto}.
\end{remark}

Apart from the well-known Borch's condition, Proposition~\ref{prop:Pareto} requires condition \eqref{eq:supergrad:cond}, which is a joint supergradient of the criterion at each point of the allocation. 
The following Proposition~\ref{prop:supergrad:com:criterion} shows that this condition is typically satisfied in case of a comonotone coherent criterion $\Ucal$ and a strictly comonotone allocation $(Z_1,\ldots,Z_n)$. 
To this end, let us recall the definition of comonotonicity:

\begin{definition}
An allocation $(Z_1,\ldots, Z_n)\in \mathbb{A}_n(X)$ is called comonotone if $Z_i=f_i(X)$ for a non-decreasing function $f_i:\R\to \R$, $i=1,\ldots,n$, and $\sum_{i=1}^nf_i=\operatorname{id}_\R$ where $\operatorname{id}_\R$ denotes the identity mapping on $\R$.

\smallskip\noindent
A comonotone allocation $(Z_1,\ldots, Z_n)\in \mathbb{A}_n(X)$ is called strictly comonotone if the above functions $f_i$, $i=1,\ldots, n$, can be chosen to be strictly increasing. 

\smallskip\noindent
A function $\Ucal:L^1\to \R$ is called comonotone if $\Ucal(\sum_{i=1}^n Z_i)=\sum_{i=1}^n \Ucal(Z_i)$ for any comonotone allocation $(Z_1,\ldots, Z_n)\in \mathbb{A}_n(X)$ for all $X\in L^1$ and $n\in \N$. 
If $\Ucal$ also satisfies conditions i)--vi), then we call $\Ucal$ a comonotone coherent criterion.
\end{definition}

Note that any linear risk sharing rule of type \eqref{eq:linearrisksharingrule} defines a comonotone allocation of $S_n$ that is strictly comonotone whenever $a_n^i>0$, $i=1,\ldots,n$. 
Indeed, $Y_i^n=f_i(S_n)$ where $f_i(x)=a_n^ix + (\E{X_i}-a_n^i\E{S_n})$, $x\in \R$, $i=1,\ldots,n$.

In the following, we briefly recall a well-known representation result for comonotone coherent criteria based on building blocks of the form % To this end let % will consider preferences given by a numerical representation $\Ucal$ constructed from building blocks of the form 
\begin{equation}
\Ucal_\lambda(X):=\frac{1}{\lambda}\int_{0}^{\lambda}q_X(t)\,\mathrm{d}t, \quad X\in L^1,\lambda\in (0,1],
\label{eq:AVaR}
\end{equation} where  the (left-continuous) quantile function $q_X$ of $X$ is given in \eqref{eq:VaR}. 
Note that $\Ucal_\lambda(X)=-\avar_\lambda(X)$ where $\avar_\lambda$ denotes the Average Value at Risk or Expected Shortfall, which is a prominent capital requirement in risk management, see, e.g.,\ F\"ollmer and Schied \cite{FS11} for a detailed discussion of the Average Value at Risk. 
We have  $\Ucal_1(\cdot)=\E{\cdot}$, and we set $\Ucal_0(X):=\essinf X$ by convention. In particular, if $\lambda\in (0,1]$, then $\Ucal_\lambda$ satisfies the conditions i)--vi) of Section~\ref{sec:pre}. 
As regards the dual representation of $\Ucal_\lambda(\cdot)$ we have (see e.g.,\ F\"ollmer and Schied \cite[Section 4.4)]{FS11})
%{\color{blue} $\Ucal_\lambda(X)=-\max_{Q\in\mathcal{Q}_{\lambda}}\EQ{-X}$}
%a dual representation of $\Ucal_\lambda(\cdot)$ as a robust expectation is 
\begin{equation}
\Ucal_{\lambda}\left(X\right)=\min_{Q\in\mathcal{Q}_{\lambda}}\EQ{X},
\label{eq:AVaRdual}
\end{equation}
where 
\begin{equation*}
\mathcal{Q}_{\lambda}:=\Qcal(\Ucal_\lambda)=\left\{Q\ll P|\frac{\mathrm{d}Q}{\mathrm{d}P}\leq \frac{1}{\lambda}\right\}, \quad \lambda\in (0,1], \quad \mathcal{Q}_{0}:=\Qcal(\Ucal_0) = \left\{Q\ll P \right\}.
\end{equation*}
Up to sign change, a comonotone coherent criterion $\Ucal$ is also known as a  {\em concave distortion risk measure} and can be represented as a mixture of building blocks \eqref{eq:AVaR} as follows (F\"ollmer and Schied \cite[Sections 4.6 and 4.7]{FS11}):
%{\color{blue} $\Ucal_\mu(X)=-\rho_{\mu}(X)=-$Choquet$[-X]$. $\Ucal_\mu(u(X))=-\rho_{\mu}(u(X))=-$Choquet$[-u(X)]$.}
\begin{equation}
\Ucal(X)=\Ucal_{\mu}\left(X\right):=\int_{[0,1]}\Ucal_{\lambda}\left(X\right)\mu(\diff \lambda).
\label{eq:Umu}
\end{equation}
Here, $\mu$ is a probability measure supported on $[0,1]$.
%We refer to Machina and Schmeidler \cite{MS92} and Ravanelli and Svindland \cite{RS14} for the link between \textit{law invariance} and \textit{probabilistic sophistication}. 

\begin{proposition}\label{prop:supergrad:com:criterion}
Suppose that $\Ucal_\mu$  satisfies $\mu(\{0\})=0$ and \begin{equation}\label{eq:mu:cond}
 \int_{(0, 1]}\frac1\lambda \mu(d\lambda)<\infty.
\end{equation} 
Then condition \eqref{eq:supergrad:cond} is satisfied for every  strictly comonotone allocation $(Z_1,\ldots, Z_n)\in \mathbb{A}_n(X)$ such that $\operatorname{Im}(Z_i)\subset \operatorname{dom} u$ for all $i=1,\ldots ,n$. 
A joint supergradient  $Q_X$ is given by the density \begin{equation*}\frac{dQ_X}{dP}:=\int_0^1\frac{1}{\lambda}(1_{\{X<q_X(\lambda)\}}+\kappa(\lambda,X) 1_{\{X=q_X(\lambda)\}})\mu(d\lambda),\end{equation*}
where \begin{equation*}\kappa(\lambda,X)=\begin{cases} 0 & \mbox{if}\, P(X=q_X(\lambda))=0 \\ \frac{\lambda-P(X<q_X(\lambda))}{P(X=q_X(\lambda))} & \mbox{if} \, P(X=q_X(\lambda))>0 \end{cases}.\end{equation*}
\end{proposition}

\begin{proof}
Step 1: Assume that $\mu=\delta_\lambda$, that is, $\Ucal_\mu(X)=\Ucal_{\lambda}$, for some $\lambda\in (0,1]$. 
For $Y\in L^1$, the probability measure $Q_Y$ given by the density $$\frac{dQ_Y}{dP}= \frac{1}{\lambda}(1_{\{Y<q_Y(\lambda)\}}+\kappa(\lambda,Y) 1_{\{Y=q_Y(\lambda)\}}),$$ where
\begin{equation}\label{eq:kappa}\kappa(\lambda,Y)=\begin{cases} 0 & \mbox{if}\, P(Y=q_Y(\lambda))=0 \\ \frac{\lambda-P(Y<q_Y(\lambda))}{P(Y=q_Y(\lambda))} & \mbox{if} \, P(Y=q_Y(\lambda))>0 \end{cases},\end{equation} satisfies $Q_Y\in \Qcal_\lambda$  and $\Ucal_\lambda(Y)=\mathrm{E}_{Q_Y}[Y]$, see for instance \cite[Remark 4.53]{FS11}. 
As $Z_i=f_i(X)$ where $f_i:\R\to \R$ is strictly increasing, and as $u\circ f_i$ is increasing on $\operatorname{dom} u$, it follows that $u(f_i (q_{X}(s)))$ is a quantile function of $u(Z_i)$ and indeed  $q_{u(Z_i)}(s)=u(f_i(q_{X}(s)))$ for all $s\in (0,1)$ outside an at most countable set, see \cite[Lemma A23]{FS11}. 
Note that $u\circ f_i$ is indeed strictly increasing on $\operatorname{dom} u$, and therefore $$\{X<q_X(\lambda)\}=\{u(Z_i)<u(f_i(q_{X}(\lambda)))\}$$ and $$\{X=q_X(\lambda)\}=\{u(Z_i)=u(f_i(q_{X}(\lambda)))\}.$$ $q_Y$ being the left-continuous quantile function implies that either $q_{u(Z_i)}(\lambda)=u(f_i(q_{X}(\lambda)))$ or $q_{u(Z_i)}(\lambda)<u(f_i(q_{X}(\lambda)))$ and $$P(u(Z_i)\in [q_{u(Z_i)}(\lambda),u(f_i(q_{X}(\lambda)))])=0.$$ 
Thus $P(X<q_X(\lambda))=P(u(Z_i)<q_{u(Z_i)})$ and $P(X=q_X(\lambda))=P(u(Z_i)=q_{u(Z_i)}(\lambda))$, and hence, $P$-a.s., $1_{\{X<q_X(\lambda)\}}=1_{\{u(Z_i)<q_{u(Z_i)}(\lambda)\}}$ and $1_{\{X=q_X(\lambda)\}}=1_{\{u(Z_i)=q_{u(Z_i)}(\lambda)\}}$. 
This implies that $Q_X=Q_{u(Z_i)}$ for all $i=1,\ldots,n$ and that $\Ucal_\lambda(u(Z_i))=\mathrm{E}_{Q_X}[u(Z_i)]$ for all $i=1,\ldots,n$. 

\smallskip\noindent
Step 2: Now consider $\Ucal_\mu$ such that $\mu(\{0\})=0$. By \eqref{eq:Umu} and Fubini, for all $Y\in L^1$,
\begin{eqnarray*}
\Ucal_\mu(Y)&=& \int_0^1 \E{\frac{1}{\lambda}(1_{\{Y<q_Y(\lambda)\}}+\kappa(\lambda,Y) 1_{\{Y=q_Y(\lambda)\}})Y} \mu(d\lambda)\\ &=& \E{\int_0^1\frac{1}{\lambda}(1_{\{Y<q_Y(\lambda)\}}+\kappa(\lambda,Y) 1_{\{Y=q_Y(\lambda)\}})\mu(d\lambda)\;Y}, 
\end{eqnarray*} where $\kappa(\lambda,Y)$ is given in \eqref{eq:kappa}.
Let \begin{equation}\label{eq:density}\varphi_X:=\int_0^1\frac{1}{\lambda}(1_{\{X<q_X(\lambda)\}}+\kappa(\lambda,X) 1_{\{X=q_X(\lambda)\}})\mu(d\lambda).\end{equation} The density $\varphi_X$ with respect to $P$ defines a probability measure $Q^\mu_X$ on $(\Omega,\mathcal{F})$ such that $\Ucal_\mu(X)=\mathrm{E}_{Q^\mu_X}[X]$. For any of the at most countable atoms $\lambda\in \{\alpha\in [0,1]\mid \mu(\{\alpha\})>0\}$ of $\mu$ (if there are any) we observe that, 
$$1_{\{X<q_X(\lambda)\}}=1_{\{u(Z_i)<q_{u(Z_i)}(\lambda)\}}\quad \mbox{and}\quad 1_{\{X=q_X(\lambda)\}}=1_{\{u(Z_i)=q_{u(Z_i)}(\lambda)\}}\quad P\mbox{-a.s.,}$$
as in step 1. 
Otherwise, the set of $\lambda\in (0,1]$ where $q_{u(Z_i)}(\lambda)<u(f_i(q_{X}(\lambda)))$ is at most countable. 
Hence, $\varphi_{u(Z_i)}=\varphi_X$, and we obtain $\Ucal_\mu(u(Z_i))=\mathrm{E}_{Q_X}[{u(Z_i)}]$ for all $i=1,\ldots,n$. 
Finally, we verify that indeed $Q^\mu_X\in \Qcal(\Ucal_\mu)$. The condition $\int_{(0, 1]}\frac1\lambda \mu(d\lambda)<\infty$ implies that $\varphi_X$ is bounded. 
Moreover, for all $Y\in L^1$, by Fubini and \eqref{eq:AVaRdual}, we have  \begin{eqnarray} \mathrm{E}_{Q_X}[{Y}]= \E{\varphi_X Y}&=& \int_0^1 \E{\frac{1}{\lambda}(1_{\{X<q_X(\lambda)\}}+\kappa(\lambda,X) 1_{\{X=q_X(\lambda)\}})Y} \mu(d\lambda) \nonumber\\ &\geq & \int_0^1 \Ucal_\lambda(Y) \mu(d\lambda) \quad = \quad \Ucal_\mu(Y).
\label{eq:QX}\end{eqnarray}
\end{proof}

Propositions~\ref{prop:Pareto} and \ref{prop:supergrad:com:criterion} imply:

\begin{corollary}\label{cor:Pareto:Umu}
Let $\Ucal=\Ucal_\mu$ as in \eqref{eq:Umu} where $\mu(\{0\})=0$ and suppose that condition \eqref{eq:mu:cond} holds. 
Further, suppose that $u$ is differentiable 
on $\operatorname{int}\operatorname{dom}u$ and that the strictly comonotone allocation $(Z_1,\ldots, Z_n)\in \mathbb{A}_n(X)$ of $X\in L^1$  satisfies Borch's condition \eqref{eq:Borch}. 
Then $(Z_1,\ldots, Z_n)$ is Pareto optimal.
\end{corollary}

\begin{corollary}\label{cor:Pareto:HARA}
Let $\Ucal=\Ucal_\mu$ as in \eqref{eq:Umu} where $\mu(\{0\})=0$ and \eqref{eq:mu:cond} is satisfied. 
Further suppose that $u$ is either a power utility, that is,  $u(x)=\frac1\gamma x^\gamma$, $x\geq 0$, (and $u(x)=-\infty$ if $x< 0$) where $\gamma\in (0,1)$, or a logarithmic utility, that is, $u(x)=\log(x)$, $x>0$, (and $u(x)=-\infty$ if $x\leq 0$). 
For $X\in L^1$ such that $P(X>0)=1$ any allocation of type  
\begin{equation}\label{eq:true:lin:rule:1}Z_1=a_1X, \ldots, Z_n=a_nX\end{equation}
for some constants $a_i>0$, $i=1,\ldots,n$, such that $\sum_{i=1}^na_i=1$ is Pareto optimal. In particular, the proportional risk sharing rule of the aggregate risk $S_n$ given by \eqref{eq:linearrisksharingrule} and $a^{i}_n=\frac{\E{X_i}}{\E{S_n}}$ is Pareto optimal.
\end{corollary}

\begin{proof}
Borch's condition reads
\begin{equation}\label{eq:HARA:Borch:1}\lambda_1 Z_1^{\gamma-1}= \ldots = \lambda_n Z_n^{\gamma-1}\end{equation} for constants $\lambda_i>0$, $i=1,\ldots, n$, where the logarithmic case corresponds to $\gamma=0$. For an allocation of type \eqref{eq:true:lin:rule:1}, \eqref{eq:HARA:Borch:1} is satisfied for $\lambda_i=(a_i)^{1-\gamma}$, $i=1,\ldots,n$. Hence, \eqref{eq:true:lin:rule:1} is Pareto optimal according to Corollary~\ref{cor:Pareto:Umu}. 
\end{proof}

The following result shows that for utilities $u$ as in Corollary~\ref{cor:Pareto:HARA} and $\Ucal(\cdot)=\E{\cdot}$, the proportional risk sharing rule is essentially the only linear risk sharing rule of type \eqref{eq:linearrisksharingrule} that defines a Pareto optimal allocation of $S_n=X_1+\ldots + X_n$ for any choice of positive initial risks $X_1,\ldots, X_n$.

\begin{lemma}
Let $\Ucal=\E{\cdot}$ (that is, $\mu=\delta_1$ in \eqref{eq:Umu}) and $u$ be as in Corollary~\ref{cor:Pareto:HARA}. 
Assume that $S_n$ is a discrete positive random variable taking at least two distinct values. 
Then the proportional risk sharing rule is the only linear risk sharing rule of type \eqref{eq:linearrisksharingrule} that defines a Pareto optimal allocation of $S_n$. %Then this risk sharing rule necessarily coincides with the proportional risk sharing rule, that is  $a^{i}_n=\frac{\E{X_i}}{\E{S_n}}$.
%\begin{equation}\label{eq:true:lin:rule}Z_1=a_1X+b_1, \ldots, Z_n=a_nX+b_n\end{equation} for some constants $a_i>0$, $b_i\in \mathbb{R}$ , $i=1,\ldots,n$, such that $\sum_{i=1}^na_i=1$, $\sum_{i=1}^n b_i=0 $, and $P(Z_i>0)=1$ for all $i=1,\ldots,n$. Then $(Z_1,\ldots, Z_n)$ is Pareto optimal if $b_i/a_i=b_1/a_1$ for all $i=1,\ldots,n$. 
%In particular a  
%linear risk sharing of type \eqref{eq:linearrisksharingrule} is Pareto optimal if and only if it coincides with the proportional risk sharing rule. 
\end{lemma}

\begin{proof} Let $(Z_1,\ldots, Z_n)\in \mathbb{A}(X)$ with $P(Z_i>0)=1$, $i=1,\ldots, n$, be a Pareto optimal allocation of the discrete positive random variable $X=\sum_{k=1}^\infty x_k1_{A_k}$ where $x_k>0$ and $P(A_k)>0$ for all $k\in \N$. Assume further that  also $Z_i=\sum_{k=1}^\infty z^i_k1_{A_k}$ for all $i=1,\ldots, n$ (which is for instance the case for any comonotone allocation, and thus for any allocation given by linear risk sharing rules). 
According to Lemma~\ref{lem:aux} below, there must exist $\lambda_1>0, \ldots, \lambda_n>0$ such that \begin{equation}\label{eq:HARA:Pareto}\sum_{i=1}^{n}\lambda_i \E{u(Z_i)}\geq \sum_{i=1}^{n}\lambda_i \E{u(Y_i)}\end{equation} for any other allocation $(Y_1,\ldots, Y_n)\in \mathbb{A}(X)$. 
By the Kuhn-Tucker theorem (\cite[Corollary 28.3.1]{Rock70}) we must have
\begin{equation}\label{eq:HARA:Borch:2} \forall k\in \N: \quad \lambda_1 (z^1_k)^{\gamma-1}= \ldots = \lambda_n (z^n_k)^{\gamma-1} \end{equation}
Indeed,  \eqref{eq:HARA:Borch:2} is a necessary and sufficient condition for $$\forall k\in \N: \quad \sum_{i=1}^n\lambda_i u(z^i_k)\geq \sum_{i=1}^n\lambda_i u(y_i) $$ for all $(y_1,\ldots, y_n)$ such that $\sum_{i=1}^ny_i=x_k$. 
If \eqref{eq:HARA:Borch:2} is not satisfied, then there is a $\hat k$ and $(y_1,\ldots, y_n)$ such that $\sum_{i=1}^ny_i=x_{\hat k}$ and $\sum_{i=1}^n\lambda_i u(z^i_{\hat k})< \sum_{i=1}^n\lambda_i u(y_i)$. 
Consider the allocation $(\tilde Z_1,\ldots, \tilde Z_n)\in \mathbb{A}(X)$ given by 
$$\tilde Z_i =\sum_{k=1, k\neq \hat k}^\infty z^i_k1_{A_k} + y_i1_{A_{\hat k}}.$$ Then \begin{eqnarray*} \sum_{i=1}^{n}\lambda_i \E{u(\tilde Z_i)} &=& \sum_{k\in \N\setminus\{\hat k\}}P(A_k)\sum_{i=1}^{n}\lambda_i  u(z^i_k) +  P(A_{\hat k})\sum_{i=1}^{n}\lambda_i  u(y_i) \\ &>& \sum_{k\in \N}P(A_k)\sum_{i=1}^{n}\lambda_i  u(z^i_k) \\&=& \sum_{i=1}^{n}\lambda_i \E{u(Z_i)},  \end{eqnarray*} a contradiction to \eqref{eq:HARA:Pareto}. 
Now suppose that in addition $Z_i=a_iX + b_i$ where $a_i>0$, $b_i\in \R$, $i=1,\ldots, n$, such that $\sum_{i=1}^na_i=1$, $\sum_{i=1}^nb_i=0$. 
By \eqref{eq:HARA:Borch:2} $$\forall i=1,\ldots, n\; \forall k\in \N\quad \lambda_i^\frac{1}{\gamma-1} (a_ix_k +b_i) = \lambda_1^\frac{1}{\gamma-1}(a_1x_k + b_1).$$
Assuming that $X$ takes at least two distinct values, the previous relation can be satisfied if and only if $\lambda_i^\frac{1}{\gamma-1} a_i=\lambda_1^\frac{1}{\gamma-1}a_1$ and $\lambda_i^\frac{1}{\gamma-1} b_i=\lambda_1^\frac{1}{\gamma-1} b_1$ for all $i=1,\ldots, n$. Recalling that $\sum_{i=1}^nb_i=0$, we conclude that $b_1=0$ and thus $b_i=0$ for all $i=1\ldots, n$. Now identify $X=S_n$ for the aggregate risk $S_n=\sum_{i=1}^nX_i$ and a risk sharing $Z_i$ of type  \eqref{eq:linearrisksharingrule}. In that case $a_i=a^i_n$ and $b_i=\E{X_i}-a^i_n\E{S_n}$. The necessary condition for Pareto optimality $b_i=0$ now implies $$a^i_n=\frac{\E{X_i}}{\E{S_n}}=a^{i, \text{prop}}_n, \quad i=1,\ldots,n.$$
\end{proof}

\begin{lemma}\label{lem:aux}
Let $\Ucal(\cdot) = \E{\cdot}$, and suppose $(Z_1,\ldots, Z_n)\in \mathbb{A}_n(X)$ is a Pareto optimal allocation of $X\in L^1$. Then the corresponding Negishi weights in \eqref{eq:Pareto:SWF} have to satisfy $\lambda_i>0$, $i=1\ldots,n$.
\end{lemma}

\begin{proof}
Suppose w.l.o.g.\ $\lambda_1=0$ and $\lambda_2>0$, then altering the allocation $(Z_{1},Z_{2}, \ldots, Z_n)\in\mathbb{A}_n(X)$ to $(\tilde Z_1, \tilde Z_2,\ldots, \tilde Z_n):=(Z_{1}-k,Z_{2}+k, \ldots, Z_n)$ for $k>0$ yields \begin{eqnarray*}\sum_{i=1}^n\lambda_i\EQ{u(\tilde Z_i)}&=& \lambda_2 \EQ{u(Z_2+k)} +\sum_{i=3}^n\lambda_i\EQ{u( Z_i)} \\&>&\lambda_2 \EQ{u(Z_2)} +\sum_{i=3}^n\lambda_i\EQ{u( Z_i)} \\ &=& \sum_{i=1}^n\lambda_i\EQ{u(Z_i)}\end{eqnarray*} since the strict monotonicity of $u$ implies that $\EQ{u(Z_2+k)}>\EQ{u(Z_2)}$.
%Hence, Negishi weights taking the value zero cannot characterize any Pareto optimum $(Z_{1},Z_{2}, \ldots, Z_n)$ in this case.
\end{proof}

%%%%%%%%%%%%%%%%%%

Finally, we generalize further to robust expectations of type 
\begin{equation}
\Ucal_{\Gamma}(X):=\inf_{\mu\in\Gamma}\Ucal_{\mu}(X)=\inf_{\mu\in\Gamma}\int_{(0,1]}\Ucal_{\lambda}\left(X\right)\mu(\diff \lambda) ,
\label{eq:lawinvariantcoh}
\end{equation}
where $\mathcal{M}_{1}((0,1])$ is the set of (Borel) probability measures on $(0,1]$ and $\Gamma\subset\mathcal{M}_{1}((0,1])$ is not empty.
From Kusuoka \cite{K01} (see also F\"ollmer and Schied  \cite[Corollary 4.63]{FS11}, Dana \cite{D05}, and Frittelli and Rosazza Gianin \cite{FRG05})
we know that, upon a sign change, all \textit{law-invariant coherent risk measures} are of the form \eqref{eq:lawinvariantcoh}.\footnote{Note that the required {\em continuity from above} in \cite[Corollary 4.63]{FS11} is automatically satisfied, see \cite{SVI10}.} %While $\eqref{eq:lawinvariantcoh}$ may be seen to account for Knightian uncertainty {\color{blue} elaborate}, by adding a strictly increasing and concave utility function $u:\R\to \R$ we incorporate a non-linear view on risk. Hence, we consider  criteria of type 
%\begin{equation}\label{eq:criterion}
%\Ucal_{\mathcal{M}}(u(X))=\inf_{\mu\in\mathcal{M}}\Ucal_{\mu}(u(X)).
%\end{equation} 
% Note that criteria of the form $\Ucal_\Gamma(u(\cdot))$ cover the popular rank-dependent utility (RDU) model (Quiggin \cite{Q82}), namely when $\Gamma=\{\mu\}$, that is in case of \eqref{eq:Umu}.
%Moreover, the RDU model encompasses Yaari's \cite{Y87} dual theory of choice under risk and the expected utility model as special cases.
The following lemma gives a sufficient condition for $L^1$-continuity of $\Ucal_\Gamma$.

\begin{lemma}\label{lem:1} Suppose that there exists $\mu \in \Gamma$ such that \eqref{eq:mu:cond} holds. 
%\begin{equation}
%\exists \mu \in \Gamma: \; \int_{(0, 1]}\frac1\lambda \mu(d\lambda)<\infty.
%\label{eq:condition:1}
%\end{equation} 
Then $\Ucal_\Gamma$ is Lipschitz continuous with respect to $L^1$-convergence.
\end{lemma}

\begin{proof}
Indeed, by superadditivity of $\Ucal_\Gamma$ $$\Ucal_\Gamma(X-Y)\leq \Ucal_\Gamma(X)-\Ucal_\Gamma(Y)\leq -\Ucal_\Gamma(Y-X). $$ 
Further, from the dual representation of $\Ucal_\lambda$ in \eqref{eq:AVaRdual} we infer that \begin{eqnarray*}\Ucal_\Gamma(X)&=& \inf_{\mu\in \Gamma}\Ucal_\mu(X) \quad = \quad  \inf_{\mu\in \Gamma}\int_{(0, 1]}\Ucal_\lambda(X) \mu(d\lambda) \\ &  \geq & \inf_{\mu\in \Gamma} \int_{(0, 1]}\min_{Q\in\mathcal{Q}_{\lambda}}\E{-\frac{dQ}{dP}|X|}  \mu(d\lambda) \quad \geq \quad   \E{-|X|}  \inf_{\mu\in \Gamma} \int_{(0, 1]} \frac{1}{\lambda} \mu(d\lambda). \end{eqnarray*} In sum we obtain $$|\Ucal_\Gamma(X)-\Ucal_\Gamma(Y)|\leq  \max\{-\Ucal_\Gamma(X-Y),-\Ucal_\Gamma(Y-X) \}\leq \E{|X-Y|} \inf_{\mu\in \Gamma}\int_{(0, 1]}\frac1\lambda \mu(d\lambda),$$ which proves the Lipschitz-continuity of $\Ucal_\Gamma$.
\end{proof}

If the condition of the previous lemma is satisfied, then $\Ucal_\Gamma$ meets the requirements i)--vi) of Section~\ref{sec:pre}, and thus $\Ucal_\Gamma=\mathrm{E}_{\Qcal(\Ucal_{\Gamma})}$. 
Under the condition that the utility $u$ is linear, the following Proposition~\ref{prop:Pareto:AVAR3} shows that any linear risk sharing \eqref{eq:linearrisksharingrule} provides a Pareto optimal allocation of $S_n$. 
Also note that this result holds in particular for criteria $\Ucal_\mu$ or $\Ucal_\lambda$, which all fall under the class $\Ucal_\Gamma$.
%In particular, $\Ucal_\Gamma$ is a law-invariant robust expectation and thus admits a representation $\Ucal_\Gamma=\mathrm{E}_\Mcal$ as in \eqref{eq:robust:exp}. 
%$\Mcal$ is not necessarily unique. 

%{\color{red} In the situation of the corollary below, we know that $\Ucal(X)=\min_{Q\in \mathcal{M}}\EQ{X}$ by continuity. Can we conclude that $\Ucal_\Gamma(Y)=\min_{\mu\in \Gamma}\Ucal_\mu (Y)$?}

\begin{proposition}\label{prop:Pareto:AVAR3}
Let $\Ucal=\Ucal_\Gamma$ as in \eqref{eq:lawinvariantcoh} and suppose that all $\mu\in \Gamma$ satisfy \eqref{eq:mu:cond}. 
Moreover, assume that $\Ucal_\Gamma(Y)=\min_{\mu\in \Gamma}\Ucal_\mu(Y)$, i.e.,\ the infimum in \eqref{eq:lawinvariantcoh} is always attained. 
If $u=\operatorname{id}_\R$ is the linear utility and if the allocation $(Z_1,\ldots, Z_n)\in \mathbb{A}_n(X)$ of $X\in L^1$ is affine, that is,  
$Z_i=a_iX+b_i$ for $a_i>0$, $b_i\in \R$, $i=1,\ldots,n$, such that $\sum_{i=1}^n a_i=1$, and $\sum_{i=1}^n b_i=0$, then $(Z_1,\ldots, Z_n)$ is Pareto optimal.
\end{proposition}

%\begin{corollary}\label{cor:Pareto:AVAR3}{\color{red} proof is incomplete}
%Let $\Ucal=\Ucal_\Gamma$ as in \eqref{eq:lawinvariantcoh} and suppose that \eqref{eq:condition:1} is satisfied. Moreover, assume that $\Ucal_\Gamma(Y)=\min_{\mu\in \Gamma}\Ucal_\mu(Y)$, i.e.\ the infimum in \eqref{eq:lawinvariantcoh} is always attained. If $u$ is differentiable 
%on $\operatorname{int}\operatorname{dom}u$ and if the allocation $(Z_1,\ldots, Z_n)\in \mathbb{A}_n(X)$ of $X\in L^1$ satisfies Borch's condition \eqref{eq:Borch} and $Z_i=f_i(X)$ for some strictly increasing functions $f_i:\R\to \R$, $i=1,\ldots, n$, such that $\sum_{i=1}^nf_i=\operatorname{id}$, then $(Z_1,\ldots, Z_n)$ is Pareto optimal.
%\end{corollary}

\begin{proof}
By assumption, 
\begin{equation}\label{eq:help:a}\Ucal_\Gamma(X)=\min_{\nu\in \Gamma}\Ucal_\nu(X) = \Ucal_{\mu_X}(X)=\mathrm{E}_{Q_X}[ X],\end{equation} where $\mu_X\in \Gamma$ and $Q_X$ is given by the density \eqref{eq:density} (with $\mu=\mu_X$). 
Note that $Q_X\in \Qcal(\Ucal_\Gamma)$ since $$\Ucal_\Gamma(Y)=\min_{\nu\in \Gamma}\Ucal_\nu(Y) \leq  \Ucal_{\mu_X}(Y)\leq \mathrm{E}_{Q_X}[ Y]$$ for all $Y\in L^1$, where the last inequality is shown in \eqref{eq:QX}. 
\eqref{eq:help:a} implies 
\begin{equation*}\mathrm{E}_{Q_X}[ X]\leq \EQ{X} \quad \mbox{for all}\; Q\in \Qcal(\Ucal_\Gamma).\end{equation*}
Hence, also $$\mathrm{E}_{Q_X}[ Z_i]=a_i\mathrm{E}_{Q_X}[X]+b_i\leq a_i\EQ{X} + b_i= \EQ{Z_i} \quad \mbox{for all}\; Q\in \Qcal(\Ucal_\Gamma).$$ The latter implies that $\Ucal_\Gamma(Z_i)=\mathrm{E}_{Q_X}[ Z_i]$ for all $i=1,\ldots, n$. 
Noting that Borch's condition is trivially satisfied when $u(x)=x$, we conclude by invoking Proposition~\ref{prop:Pareto}.
%
%From the proof of Corollary~\ref{cor:Pareto:AVAR2} we also know that $\Ucal_{\mu_X}(u(Z_i))=\mathrm{E}_{Q_X}[ u(Z_i)]$ for all $i=1,\ldots,n$. Therefore, the assertion follows from Proposition~\ref{prop:Pareto} once we show that $\Ucal_\Gamma(u(Z_i))=\Ucal_{\mu_X}(u(Z_i))$ for all $i=1,\ldots,n$.
%In particular, we have 
%\begin{equation}\label{eq:help:1}\mathrm{E}_{Q_X}[ X]\leq \EQ{X} \quad \mbox{for all}\; Q\in \mathcal{M}.\end{equation} We show that $\mathrm{E}_{Q_X}[ u(f_i(X))]\leq \EQ{u(f_i(X))}$ for all $i=1,\ldots,n$ and all $Q\in \mathcal{M}$. This then implies $\mathrm{E}_{Q_X}[ u(f_i(X))]= \Ucal(u(f_i(X)))$ for all $i=1,\ldots,n$ and thus the assertion follows as in the proof of the previous Corollaries~\ref{cor:Pareto:AVAR} and \ref{cor:Pareto:AVAR2}.
%
%Suppose that $x\mapsto ax+b$ is an affine majorant of the concave and increasing function $u\circ f_i$, $i=1,\ldots,n$. Note that any such affine majorant satisfies $a\geq 0$ since $u\circ f_i$ is increasing. Hence, by \eqref{eq:help:1} we obtain for arbitrary $Q\in \mathcal{M}$ $$\mathrm{E}_{Q_X}[ u(f_i(X))]\leq \mathrm{E}_{Q_X}[ aX+b]\leq \EQ{aX+b}.$$ Taking the infimum over all such affine majorants yields $$\mathrm{E}_{Q_X}[ u(f_i(X))]\leq \EQ{u(f_i(X))}.$$ As $Q\in \mathcal{M}$ was arbitrary, taking the infimum over all $Q\in \mathcal{M}$ we obtain $\mathrm{E}_{Q_X}[ u(f_i(X))]\leq \Ucal(u(f_i(X))),$ and indeed $\mathrm{E}_{Q_X}[ u(f_i(X))]= \Ucal(u(f_i(X)))$ since $Q_X\in \mathcal{M}$.
\end{proof}

 %Proposition~\ref{prop:Pareto:AVAR3} further reveals how Pareto optimality depends on the utility function $u$. Indeed, the following example in particular shows that under the same conditions for $\Ucal$ as in Proposition~\ref{prop:Pareto:AVAR3}, replacing the linear utility by the exponential, linear risk sharing, apart from the case when it coincides with the equal risk sharing rule, is not optimal any longer.

\begin{example}
For exponential utility $u(x)=1-\exp(-\gamma x)$, $\gamma>0$, $x\in\R$, Borch's condition (\ref{eq:Borch})  takes the form
$$\lambda_1 \exp(-\gamma Z_1)= \ldots = \lambda_n\exp(-\gamma Z_n)$$ for constants $\lambda_i>0$, $i=1,\ldots, n$. This is clearly satisfied for the equal risk sharing rule, but does not hold for  other affine allocations (or linear risk sharing rules) unless $X$ is constant.
\end{example}

%\begin{example}
%Consider the linear utility $u(x)=x$, $x\in\R$. Then Borch's condition (\ref{eq:Borch}) is always satisfied. Hence, any strictly comonotone allocation is Pareto optimal in case $\Ucal=\Ucal_\mu$ satisfies the conditions of Corollary~\ref{cor:Pareto:Umu}. This well-known result is also easily directly verified since $\Ucal_\mu$ is {\em comonotonic additive}, see F\"ollmer and Schied \cite[Theorem 4.93]{FS11} for the details. 
%\end{example}

\section{Applications}\label{sec:applications}

\subsection{Actuarial Pricing in Tariff Cells} Insurance companies use certain characteristics to decompose their portfolio of insurance policies into homogeneous groups, so-called tariff cells.  
If the number of policies within a tariff cell is large, then the premium of the idiosyncratic risks is derived from the law of large numbers as the expected claims amount per policy plus a suitable safety loading. 
Note that in this paper the random variables $X_1,X_2,\ldots$ model payoffs---that is, minus risks---so that $X_1,X_2,\ldots$ corresponds to the insurer\rq s payoffs resulting from a portfolio of insurance contracts. 
If the insurance company uses a utility-based premium calculation for the tariff cell, that is, charging minus the certainty equivalent $-u^{-1}(\Ucal(u(Y^{n}_i)))$ of policyholder $i$'s share of the pool, where $(Y^n_1,\ldots, Y^n_n)$ is a linear risk sharing of type \eqref{eq:linearrisksharingrule}, then the implied safety loading equals the risk premium $\pi(Y_i^n)$ since by \eqref{eq:riskpremium} $$\E{-Y^n_i}+\pi(Y^n_i)=-u^{-1}(\Ucal(u(Y^{n}_i))).$$ 
Note that, in the above expression, $\E{-Y^n_i}$ is the expected loss of policyholder $i$'s share of the pool, whereas $-u^{-1}(\Ucal(u(Y^{n}_i)))$ is the target premium. 

%{\color{red} here we should mention the risk premium somewhere}. 
The tariff cells are in general not completely homogeneous since there is a trade-off between homogeneity and a sufficiently large number of contracts per tariff cell, i.e., the portfolio decomposition relies on some main characteristics only, while other characteristics are not taken into account.
Let us assume partial inhomogeneity within a tariff cell in the sense that the means $\E{X_i}$ and variances $\var{X_i}$, $i\in\N$, slightly fluctuate around the mean and the variance of a reference risk, say $X_1$, i.e., there exist constants $\epsilon,\delta>0$ such that
$$\E{X_i}\in[\E{X_1}-\epsilon,\E{X_1}+\epsilon],\quad \var{X_i}\in[\var{X_1}-\delta,\var{X_1}+\delta].$$
In this case, we obtain---depending on the linear risk sharing rule---lower and upper bounds $c_i^{\cdot,\text{lb}}$ and $c_i^{\cdot,\text{ub}}$ for $\sqrt{n}a^{i,\cdot}_n\sigma_n$ that according to Theorem~\ref{thm:main:1} provide lower and upper bounds for the decay of the risk premium:
%\begin{eqnarray}
%\liminf_{n\rightarrow\infty}\sqrt{n}\;\pi(\E{X_i}+a_n^{i,\cdot}(S_n-\E{S_n})) 
%&\geq&-c^{\cdot,\text{lb}}_i \,\Ucal \left(Z\right)\label{eq:lb1},\\
%\limsup_{n\rightarrow\infty}\sqrt{n}\;\pi(\E{X_i}+a_n^{i,\cdot}(S_n-\E{S_n})) 
%&\leq&-c^{\cdot,\text{ub}}_i \,\Ucal \left(Z\right)\label{eq:ub1}.
%\end{eqnarray}
More precisely, we identify the following bounds, assuming that $\E{X_1}+\epsilon\leq 0$:
\begin{itemize}
    \item \emph{Proportional rule:} For
    $a^{i,\rm{prop}}_n:=\frac{\E{X_i}}{\E{S_n}}$ the term 
    $$\sqrt{n}a^{i,\rm{prop}}_n\sigma_n=\frac{\sqrt{n\sum_{j=1}^n\var{X_j}}}{\sum_{j=1}^n\E{X_j}}\E{X_i} $$
    satisfies 
    \begin{eqnarray*}
     \sqrt{n}a^{i,\rm{prop}}_n\sigma_n&\geq&\frac{\sqrt{n\cdot n(\var{X_1}-\delta)}}{n(\E{X_1}-\epsilon)}\E{X_i}=\frac{\sqrt{\var{X_1}-\delta}}{\E{X_1}-\epsilon}\E{X_i}=:c_i^{\rm{prop,lb}},\\
     \sqrt{n}a^{i,\rm{prop}}_n\sigma_n&\leq&\frac{\sqrt{n\cdot n(\var{X_1}+\delta)}}{n(\E{X_1}+\epsilon)}\E{X_i}=\frac{\sqrt{\var{X_1}+\delta}}{\E{X_1}+\epsilon}\E{X_i}=:c_i^{\rm{prop,ub}}.
     \end{eqnarray*}
    \item \emph{Linear regression rule:} If $\sqrt{n}a^{i,\rm{reg}}_n\sigma_n=\frac{\sqrt{n}}{\sigma_n}\var{X_i}$, hence
    $$c_i^{\rm{reg,lb}}:=\frac{\var{X_i}}{\sqrt{\var{X_1}+\delta}}\leq \sqrt{n}a^{i,\text{reg}}_n\sigma_n\leq \frac{\var{X_i}}{\sqrt{\var{X_1}-\delta}}:=c_i^{\rm{reg,ub}}.$$
    \item \emph{Mean-variance rule:} For
    $a^{i,\rm{mv}}_n:=\beta a^{i,\rm{prop}}_n+(1-\beta)a^{i,\rm{reg}}_n$, $\beta\in(0,1)$, the bounds take the form
    $$c_i^{\rm{mv,lb}}:=\beta c_i^{\rm{prop,lb}}+(1-\beta)c_i^{\rm{reg,lb}},\quad c_i^{\rm{mv,ub}}:=\beta c_i^{\rm{prop,ub}}+(1-\beta)c_i^{\rm{reg,ub}}.$$
\end{itemize}

\subsection{Actuarial Pricing with Deductibles}
To motivate policyholders to prevent damage or to bear minor damages themselves, insurance contracts often include deductibles. Policyholders can usually choose the amount of the deductible and thus reduce their premium by covering a part of the damage. Suppose that the original damages of the policyholders are described by the non-negative i.i.d.\,random variables $C_1,C_2,\ldots$. If each policyholder can choose an individual deductible $d_i\in[0,d^*]$, then the insurance company must only bear the claims amount $(C_i-d_i)^+$ per policyholder $i\in\mathbb{N}$. In our setting, this corresponds to the financial positions $X_i=-(C_i-d_i)^+$, $i\in\mathbb{N}$. Note that $X_1,X_2,\ldots$ is still a sequence of independent random variables, but in general subject to heterogeneity due to the different deductibles.

An analogous situation arises if we interpret $d_i$, $i\in\mathbb{N}$, as thresholds of different stop-loss reinsurance contracts for independent underlying losses $C_1,C_2,\ldots$.

We assume that $P(C_1>d^*)>0$ such that $\E{X_i}=\E{-(C_1-d^*)^+}<0$. Since $d_i\mapsto -(c-d_i)^+$ is non-decreasing, we obtain the estimate $\E{X_i}=\E{-(C_1-d_i)^+}\in [\E{-C_1},\E{-(C_1-d^*)^+}]$. Moreover, $\var{X_i}=\var{-(C_1-d_i)^+}\in[\sigma^2_{\min},\sigma^2_{\max}]$ for constants $0<\sigma^2_{\min}\leq\sigma^2_{\max}$.
Theorem~\ref{thm:main:1} provides lower and upper bounds for the decay of the risk premium:
\begin{itemize}
    \item \emph{Proportional rule:} For
    $a^{i,\rm{prop}}_n:=\frac{\E{X_i}}{\E{S_n}}$ the term 
    $$\sqrt{n}a^{i,\rm{prop}}_n\sigma_n=\frac{\sqrt{n\sum_{j=1}^n\var{X_j}}}{\sum_{j=1}^n\E{X_j}}\E{X_i} $$
    satisfies 
\begin{eqnarray*}
     \sqrt{n}a^{i,\rm{prop}}_n\sigma_n&\geq&\frac{\sqrt{n\cdot n \cdot \sigma^2_{\min}}}{n\cdot\E{-C_1}}\E{-(C_1-d_i)^+}=\frac{\E{(C_1-d_i)^+}}{\E{C_1}}\sigma_{\min}=:c_i^{\rm{prop,lb}},\\
     \sqrt{n}a^{i,\rm{prop}}_n\sigma_n&\leq&\frac{\sqrt{n\cdot n\cdot \sigma^2_{\max}}}{n\cdot\E{-(C_1-d^*)^+}}\E{-(C_1-d_i)^+}=\frac{\E{(C_1-d_i)^+}}{\E{(C_1-d^*)^+}}\sigma_{\max}=:c_i^{\rm{prop,ub}}.
     \end{eqnarray*}
     Note that
     $$\frac{\E{(C_1-d^*)^+}}{\E{C_1}}\sigma_{\min}\leq c_i^{\rm{prop,lb}}\leq c_i^{\rm{prop,ub}}\leq \frac{\E{C_1}}{\E{(C_1-d^*)^+}}\sigma_{\max}, \quad i\in\mathbb{N},$$
     i.\,e., we even obtain uniform bounds for all deductibles in the interval $[0,d^*]$.

    \item \emph{Linear regression rule:} If $\sqrt{n}a^{i,\rm{reg}}_n\sigma_n=\frac{\sqrt{n}}{\sigma_n}\var{X_i}$, hence
    $$\tfrac{\sigma^2_{\min}}{\sigma_{\max}}\leq c_i^{\rm{reg,lb}}:=\frac{\var{X_i}}{\sigma_{\max}}\leq \sqrt{n}a^{i,\text{reg}}_n\sigma_n\leq \frac{\var{X_i}}{\sigma_{\min}}:=c_i^{\rm{reg,ub}}\leq \tfrac{\sigma^2_{\max}}{\sigma_{\min}}.$$
    \item \emph{Mean-variance rule:} For
    $a^{i,\rm{mv}}_n:=\beta a^{i,\rm{prop}}_n+(1-\beta)a^{i,\rm{reg}}_n$, $\beta\in(0,1)$, the lower and upper bounds take the form
    $$c_i^{\rm{mv,lb}}:=\beta c_i^{\rm{prop,lb}}+(1-\beta)c_i^{\rm{reg,lb}},\quad c_i^{\rm{mv,ub}}:=\beta c_i^{\rm{prop,ub}}+(1-\beta)c_i^{\rm{reg,ub}},$$ respectively.
\end{itemize}
\subsection{Bounds for Rate of Convergence for Concave $\Ucal$}
 Suppose that the monotone criterion $\tilde{\Ucal}:L^1\rightarrow\R$ instead of being superadditive and positively homogeneous is only concave, that is, $\tilde \Ucal(\lambda X+ (1-\lambda)Y)\geq \lambda \tilde \Ucal (X) + (1-\lambda)\tilde \Ucal(Y)$ for all $\lambda\in [0,1]$ and $X,Y\in L^1$. 
Then, $\tilde{\Ucal}$ admits a robust representation
$$\tilde{\Ucal}(X)=\inf_{Q\in\Mcal}\left(\EQ{X}+\beta(Q)\right), $$
where $\Mcal$ is a set of probability measures on $(\Omega,\Fcal)$ as in \eqref{eq:robust:exp} and  $\beta: \Mcal\rightarrow[0,\infty]$ is a penalty function, see e.g.,\ \cite{BKMS21}. 
Denoting by 
$\Ucal={\rm E}_\mathcal{M}$ 
the corresponding coherent criterion, we obtain the estimate $\Ucal(X)\leq \tilde{\Ucal}(X)$ for all $X\in L^1$. Thus, the risk premium 
$$\tilde{\pi}(\E{X_i}+a^{i}_n(S_n-\E{S_n}))=\E{X_i}-u^{-1}(\tilde{\Ucal}(u(\E{X_i}+a^{i}_n(S_n-\E{S_n})))) $$
with respect to $\tilde{\Ucal}$ satisfies $\tilde{\pi}\leq \pi$. 
Together with Theorem~\ref{thm:main:1} this yields an upper bound for the rate of convergence
$$\limsup_{i\leq n\to\infty} \sqrt{n}\tilde{\pi}(\E{X_i}+a^{i}_n(S_n-\E{S_n}))\leq -c_i\Ucal(Z),$$
where $Z$ is a standard normal random variable with respect to the reference measure $P$.

%\subsection{Risk Sharing with Distortion Risk Measures}

%\begin{equation*}
%h_{m}(p)=\Phi\left(\Phi^{-1}(p)+m\right),\qquad m\in\mathbb{R},
%\end{equation*}

%\subsection{KMM preferences}
%{\color{red}
%Recent research (see, e.\,g., \cite{CVW25}) analyzes portfolio optimization with respect to smooth ambiguity preferences introduced in \cite{KMM05}, therefore 
%also called KMM preferences. These preferences take the form
%$$\mathcal{K}(X)=\sum_{i=1}^mw_i\phi(\mathrm{E}_{Q_i}[u(X)]),$$
%where $Q_1,\ldots,Q_m$, $m\in\mathbb{N}$ denote probability measures with specific weights $w_1,\ldots,w_m$, $u$ is a utility function, and where the function $\phi$ is strictly increasing and strictly concave. The function $\phi$ is called the ambiguity (averse) attitude. The key idea behind KMM preferences is twofold: On the one hand, the investor applies classical expected utility $\mathbb{E}_{Q_i}[u(X)]$ and weights the expected the expected utilities among all measures $Q_i$, $i=1,\ldots,m$, i.\,e., the weights $w_i$, $i=1,\ldots,m$ describe the investor's confidence
%in the probability measures $Q_i$. On the other hand, each expected utility is distorted by the investor’s ambiguity attitude $\phi$. This approach provides a distinction between ambiguity beliefs described in terms of the probability measures $Q_1,\ldots,Q_m$ and ambiguity
%attitude.}

%%%%%%%%%%%%%%%%%%%%%%%%%%%%%%%%%%%%%%%%%%%%%%%%%%%%%%%%%%%%%%%%%%%%%%%%%%

\newpage

\setcounter{equation}{0}

\begin{appendix}

\section*{APPENDIX}
\section{Proof of Theorem~\ref{thm:main:1}}\label{sec:proof:mainthm}
\begin{lemma}\label{lem:above} In the setting of Theorem~\ref{thm:main:1}, we have that 
$\Ucal\left(S_{n}^{*}\right)\rightarrow \Ucal\left(Z\right)$, where $Z$ is a standard normal random variable.
\end{lemma}

\begin{proof} 
Let $U$ be a uniformly distributed random variable on $(0,1)$, and denote by $q_n$ any quantile function of $S^\ast_n$, for instance the left-continuous one given in \eqref{eq:VaR}. Then $q_n(U)$ and $S^\ast_n$ are identically distributed, $n\in \mathbb N$, and also $\Phi^{-1}(U)$ and $Z$ are identically distributed, where $\Phi$ denotes the cumulative distribution function of $Z$. Since $S^\ast_n$ is assumed to converge in distribution to $Z$, the corresponding quantile functions converge pointwise. Moreover,  
by H\"older's inequality $$\E{|q_n(U)-\Phi^{-1}(U)|^2}\leq  \E{(S^\ast_n)^2} + \E{Z^2} + 2| \E{q_n(U)\Phi^{-1}(U)}|\leq  1+1+ 2=4.$$ Thus $(q_n(U)-\Phi^{-1}(U))_{n\in \mathbb N}$ is uniformly integrable by the de Vall\'ee-Poussin theorem. Lebesgue's convergence theorem now implies $\E{|q_n(U)-\Phi^{-1}(U)|}\to 0$, that is, $q_n(U)$ converges in $L^1$ to $\Phi^{-1}(U)$. 
Therefore, and by $L^1$-continuity and law-invariance of $\Ucal$,
$$\lim_{n\to \infty}\Ucal(S^\ast_n) = \lim_{n\to \infty}\Ucal(q_n(U))= \Ucal(\Phi^{-1}(U))= \Ucal(Z).$$
\end{proof}

\begin{proof}\textit{of Theorem~\ref{thm:main:1}}
%We may assume that $\sigma_{X_{1}}>0$ {\color{blue} (and thus $\sigma_{S_n}>0)$} since the assertion is trivial otherwise.
We start by computing a Taylor expansion of $u(\E{X_i}+a_n^i(S_n-\E{S_n}))$ around $\E{X_i}$ up to the first order:
\begin{equation}
u(\E{X_i}+a_n^i(S_n-\E{S_n}))=u(\E{X_i})+u'(Y_{n,i})a_n^i(S_n-\E{S_n}).
\label{eq:Taylor1st}
\end{equation}
Here, $Y_{n,i}$ is a random variable taking values between $\E{X_i}+a_n^i(S_n-\E{S_n})$ and $\E{X_i}$. 
Note that the requirement $| u'(\E{X_i}+a^i_n(S_n-\E{S_n}))| \leq G_i$ for all $n\in \N$ where $G_i\in L^2$ implies that there exists an $H_i\in L^2$ such that $u'(Y_{n,i})\leq H_i$ for all $n\in \N$, because $u'$ is non-negative and decreasing on $\operatorname{int}\operatorname{dom} u$ and hence $$0\leq u'(Y_n,i)\leq \max\{u'(\E{X_i}), u'(\E{X_i} +a^i_n(S_n-\E{S_n})))\}.$$ 
In particular, by \eqref{eq:Taylor1st} and H\"older's inequality it also follows that $u(\E{X_i}+a_n^i(S_n-\E{S_n}))\in L^1$ for all $n\in \N$. 

Next, by cash-additivity of $\Ucal$ and invoking the first-order Taylor expansion of $u^{-1}$,
\begin{eqnarray*}
u^{-1}\left(\Ucal\left(u\left(\E{X_i}+a_n^i(S_n-\E{S_n})\right)\right)\right) 
&=&u^{-1}\left(\Ucal\left(u(\E{X_i})+u'(Y_{n,i})a_n^i(S_n-\E{S_n})\right)\right) \\
&=&u^{-1}\left(u\left(\E{X_i}\right)+\Ucal\left(u'(Y_{n,i})a_n^i(S_n-\E{S_n})\right)\right)\\
&=&u^{-1}\circ u\left(\E{X_i}\right)\\
&&+\left(u^{-1}\right)'\left(y_{n}\right)\Ucal\left(u'(Y_{n,i})a_n^i(S_n-\E{S_n})\right)\\
&=&\E{X_i}+\left(u^{-1}\right)'\left(y_{n}\right)\Ucal\left(u'(Y_{n,i})a_n^i(S_n-\E{S_n})\right),
\end{eqnarray*}
with $y_{n}\in [u\left(\E{X_i}\right),u\left(\E{X_i}\right)+\Ucal\left(u'(Y_{n,i})a_n^i(S_n-\E{S_n})\right)]$.
Hence, by positive homogeneity of $\Ucal$
\begin{eqnarray*} 
\sqrt{n} \pi(\E{X_i}+a_n^i(S_n-\E{S_n}))&=&-\sqrt{n}\left(u^{-1}\right)'\left(y_{n}\right)\Ucal\left(u'(Y_{n,i})a_n^i(S_n-\E{S_n})\right)\nonumber\\
&=&-\sqrt{n} \sigma_n a_n^i \left(u^{-1}\right)'\left(y_{n}\right)\Ucal\left(u'(Y_{n,i})S^\ast_n\right).\label{eq:prelimit}
\end{eqnarray*}
According to Lemma~\ref{lem:above},
\begin{equation}
\label{eq:CLT}
\Ucal\left(S_{n}^{*}\right)\rightarrow \Ucal\left(Z\right),
\end{equation}
where $Z$ is a standard normal random variable. %Hence, by positive homogeneity, recalling that $u'$ takes positive values:
%\begin{equation*}
%\Ucal_{\mu}\left(u'(\E{X_1})S_{n}^{*}\right)\rightarrow \Ucal_{\mu}\left(u'(\E{X_1})Z\right).
%\end{equation*}
By superadditivity,
\begin{equation*}
\Ucal(u'(Y_{n,i})S_{n}^{*})-\Ucal(u'(\E{X_i})S_{n}^{*})\geq \Ucal((u'(Y_{n,i})-u'(\E{X_i}))S_n^*).
\end{equation*}
As $\frac{1}{n}(S_n-\E{S_n})\to 0$ $P$-a.s.\ and $\sup_{n\in \N}a_n^in<\infty$ by assumption, we have $a_n^i(S_n-\E{S_n})\to 0$ $P$-a.s.\  and thus $Y_{n,i}\to \E{X_i}$ $P$-a.s.\ for $n\to \infty$. 
Now H\"older, continuity of $u'$, and the dominated convergence theorem (recall that $u'(Y_{n,i})\leq H_i\in L^2$ for all $n\in \N$) imply
\begin{eqnarray*}\E{|(u'(Y_{n,i})-u'(\E{X_i}))S_n^*|} &\leq & \E{(S_n^*)^2}^\frac12 \E{(u'(Y_{n,i})-u'(\E{X_i}))^2}^\frac12 \\ &= & \E{(u'(Y_{n,i})-u'(\E{X_i}))^2}^\frac12 \to 0.
\end{eqnarray*}
Hence, $L^1$-continuity of $\Ucal$ yields $ \Ucal((u'(Y_{n,i})-u'(\E{X_i}))S_n^*)\to 0$, and similarly,
\begin{equation*}
\Ucal(u'(\E{X_i})S_{n}^{*})-\Ucal(u'(Y_{n,i})S_{n}^{*})\geq \Ucal((u'(\E{X_i})-u'(Y_{n,i}))S_n^*) \to 0.
\end{equation*}
Therefore, and by positive homogeneity and \eqref{eq:CLT},
\begin{equation*}
\lim_{n\to \infty}\Ucal\left(u'\left(Y_{n,i}\right)
S_{n}^{*}\right)= \lim_{n\to \infty} u'(\E{X_i})\Ucal\left(S_{n}^{*}\right) = u'(\E{X_i})\Ucal(Z).
\end{equation*}
Finally, we show that $y_n\to u(\E{X_i})$  and hence \begin{equation*}
\left(u^{-1}\right)'\left(y_{n}\right)
=\frac{1}{u'\left(u^{-1}\left(y_{n}\right)\right)}\to \frac{1}{u'\left(\E{X_i}\right)}.
\end{equation*}
Indeed, another application of H\"older's inequality yields  
$$\E{|u'(Y_{n,i})a_n^i(S_n-\E{S_n})|}\leq \E{H_i^2}^{\frac12}a^i_n\E{\left(S_{n}-\E{S_n}\right)^2}^{\frac12}\leq \E{H_i^2}^{\frac12} a^i_n\sigma_n\to 0,$$ since $\limsup_{n\to \infty}\sqrt{n}a_n^i\sigma_n<\infty$ implies $\lim_{n\to \infty}a_n^i\sigma_n=0$. Hence, $L^1$-continuity of $\Ucal$ yields $\Ucal\left(u'\left(Y_{n,i}\right)a_n^i(S_n-\E{S_n})\right)\to 0$, which implies that $y_n\to u(\E{X_i})$. 
Thus we obtain \eqref{eq:conv:rate} noting that $-\Ucal(Z)\geq 0$ according to Proposition~\ref{prop:risk:aversion}, and eventually also \eqref{eq:conv:rate:lim}.

The proof of \eqref{eq:conv:rate:no:amb} is a generalization of the proof of \cite[Theorem 2.5]{KLS16}. 
This time we compute the Taylor expansion of $u$ of order 2 around $\E{X_i}$:
\begin{eqnarray*}u\left(\E{X_i}+a_n^i(S_n-\E{S_n})\right) &=& u(\E{X_i})+u'(\E{X_i})a_n^i\left(S_n-\E{S_n}\right)+ \nonumber \\ \label{eq1} && + \frac12 u''(Z_{n,i})\left(a_n^i(S_n-\E{S_n})\right)^2,\end{eqnarray*}
where $Z_{n,i}$ takes values between $\E{X_i}$ and $\E{X_i}+a_n^i(S_n-\E{S_n})$.
Taking expectations we obtain 
\begin{equation*}\label{eq2}\E{u\left(\E{X_i}+a_n^i(S_n-\E{S_n})\right)}=u(\E{X_i})+0+\frac12 \E{u''(Z_{n,i})\left(a_n^i(S_n-\E{S_n})\right)^2}.\end{equation*}
Next we take the Taylor expansion of $u^{-1}$ around the point $u(\E{X_i})$ to obtain
\begin{align} & u^{-1}\left(\E{u\left(\E{X_i}+a_n^i(S_n-\E{S_n})\right)}\right)\nonumber \\
=&u^{-1}\left(u(\E{X_i})+\frac12 \E{u''(Z_{n,i})\left(a_n^i(S_n-\E{S_n})\right)^2}\right)\nonumber \\
=&u^{-1}\circ u(\E{X_i})+(u^{-1})'(z_{n,i})\frac12 \E{u''(Z_{n,i})\left(a_n^i(S_n-\E{S_n})\right)^2}\nonumber  \\
=&\E{X_i}+(u^{-1})'(z_{n,i})\frac{(a_n^i)^2u''(\E{X_i})\sigma^2_n}{2} \nonumber\\
& \quad+ (u^{-1})'(z_{n,i})\frac12 \E{(u''(Z_{n,i})-u''(\E{X_i}))\left(a_n^i(S_n-\E{S_n})\right)^2},  \label{eq3}\end{align}
where $z_{n,i}\in \left[u(\E{X_i})+\E{u''(Z_{n,i})\left(a_n^i(S_n-\E{S_n})\right)^2}, u(\E{X_i})\right]$.
Hence, \[n\pi(Y_i^n)= - n(u^{-1})'(z_{n,i})\frac{(a_n^i)^2u''(\E{X_i})\sigma^2_n}{2} - \Delta_{i,n}\]
where $$\Delta_{i,n}:=(u^{-1})'(z_{n,i})\frac{n}{2} \E{(u''(Z_{n,i})-u''(\E{X_i}))\left(a_n^i(S_n-\E{S_n})\right)^2}.$$
By assumption we have $\limsup_{n\to \infty} na_{n}^i\sigma^2=c_i^2$. Moreover, as in this case we assume that the $X_i$ are uniformly bounded and $\sup_{n\in \N}na_n^i<\infty$, it follows that $Z_{n,i}$ is bounded. Therefore, $$\E{u''(Z_{n,i})\left(a_n^i(S_n-\E{S_n})\right)^2}\leq K (a_n^i)^2\sigma^2_n,$$ where $K>0$ is a constant and $(a_n^i)^2\sigma^2_n$. It follows that $z_{n,i}\to \E{X_i}$ and thus
$$(u^{-1})'(z_{n,i})=\frac{1}{u'(u^{-1}(z_{n,i}))}\to \frac{1}{u'(\E{X_i})}. $$ It remains to prove that $\Delta_{n,i}\to 0$. As the $X_{i}$ are uniformly bounded, $(S_n-\E{S_n})/n$ is uniformly bounded (for all $n\in \N$) by the same constant. Hence, there is a constant $K_1>0$ such that $|a_n^i(S_n-\E{X_i})|\leq K_1$, because $\sup_{n\in \N}na_n^i<\infty$. By H\"older's inequality \begin{eqnarray*}|\Delta_{n,i}|&\leq& (u^{-1})'(z_{n,i})\frac{n}{2}\E{(u''(Z_{n,i})-u''(\E{X_i}))^2}^{\frac12}\E{(a_n^i(S_n-\E{S_n})^4}^{\frac12}\\ & \leq& (u^{-1})'(z_{n,i})\frac{n}{2}\E{(u''(Z_{n,i})-u''(\E{X_i}))^2}^{\frac12} \E{(a_n^i(S_n-\E{S_n})^2}^{\frac12}K_1 \\&=& (u^{-1})'(z_{n,i}) \frac{K_1n}{2} (a^i_n)^2\sigma_n^2 \E{(u''(Z_{n,i})-u''(\E{X_i}))^2}^{\frac12}.\end{eqnarray*} $(u^{-1})'(z_{n,i})$ is convergent, thus bounded, $\limsup_{n\to \infty}n(a_n^i)^2\sigma^2_n= c^2_i$, whereas $$\E{(u''(Z_{n,i})-u''(\E{X_i}))^2}\to 0$$ by dominated convergence. 
\end{proof}

\setcounter{equation}{0}

\section{Pareto Optimal Allocations and Borch's Condition}\label{app:Pareto}

We recall that a function $\Ucal:L^1\to \R\cup\{-\infty\}$ is {\it upper semi-continuous} whenever 
its upper level sets $L_k:=\{X\in L^1\mid \Ucal(X)\geq k\}$ are closed in $(L^1,\E{|\cdot|})$ for all $k\in \R$.

\begin{proposition} \label{prop:generalutility}Suppose that $\Ucal_i:L^1\to \R\cup\{-\infty\}$ are monotone, concave, and upper semi-continuous functions, $i=1,\ldots,n$. Moreover, suppose that $u_1,\ldots, u_n:\R\to\R\cup\{-\infty\}$, $i=1,\ldots,n$, are increasing and strictly increasing and differentiable on the interior of the respective domains $\operatorname{int}\operatorname{dom}u_i$, $i=1,\ldots,n$. Let $X\in L^1$ and suppose that the allocation $(X_1,\ldots, X_n)\in \mathbb{A}_n(X)$ satisfies Borch's condition: 
\begin{equation}\label{eq:Borch2}
    \mbox{there exist}\; \lambda_1,\ldots, \lambda_n>0 \; \mbox{such that }\; \lambda_1 u_1'(X_1)=\ldots = \lambda_n u_n'(X_n).
\end{equation}
 Moreover, suppose that there is a probability measure $Z\in \bigcap_{i=1}^n\dom\Ucal_i^\ast$ such that  $$\Ucal_i(u_i(X_i))=\E{Zu_i(X_i)}-\Ucal_i^\ast(Z)$$ for all $i=1,\ldots,n$. 
 Then $(X_1,\ldots, X_n)$ is Pareto optimal. 
 Here, $$\Ucal_i^\ast(\tilde Z)=\inf_{X\in L^1}\E{\tilde ZX}-\Ucal(X), \quad \tilde Z\in L^\infty,$$ denotes the dual function of $\Ucal_i$, $i=1\ldots,n$.
\end{proposition}

\begin{proof} Note that by the Fenchel-Moreau theorem, \cite[Part One, Proposition 3.1]{eketem99} we have for all $i=1,\ldots, n$
\begin{equation}\label{eq:dual:rep}\Ucal_i(Y)= \inf_{\tilde Z\in L^\infty} \E{\tilde ZY}-\Ucal_i^\ast(\tilde Z).\end{equation}
The Kuhn-Tucker theorem \cite[Corollary 28.3.1]{Rock70} %(applied state-wise with state-dependent Lagrange multiplier  $\mu= \lambda_1 u_1'(X_1)$) 
ensures that $$\sum_{i=1}^n\lambda_iu_i(X_i)\geq \sum_{i=1}^n\lambda_iu_i(Y_i)$$ for all $(Y_1,\ldots,Y_n)\in \mathbb{A}_n(X)$. By monotonicity of the $\Ucal_i$ we have that $Z\geq 0$ and therefore also $$\sum_{i=1}^n\lambda_i \E{Zu_i(X_i)}\geq \sum_{i=1}^n\lambda_i \E{Zu_i(Y_i)}.$$ The condition on $Z$ and \eqref{eq:dual:rep} yield 
$$\sum_{i=1}^n\lambda_i \Ucal_i(u_i(X_i))= \sum_{i=1}^n\lambda_i (\E{Zu_i(X_i)}-\Ucal^\ast_i(Z))\geq \sum_{i=1}^n\lambda_i (\E{Zu_i(Y_i)}-\Ucal^\ast_i(Z))\geq \sum_{i=1}^n\lambda_i \Ucal_i(u_i(Y_i)).$$ Hence, Pareto optimality of $(X_1,\ldots, X_n)$ follows.
\end{proof}

%\begin{comment}

The proof of the following result is a straightforward generalization of the proof of Corollary~\ref{cor:Pareto:Umu}.
\begin{corollary}
Let $\Ucal_i=\Ucal_\mu$, $i=1,\ldots,n$, for $\mu\in \Mcal_1((0,1])$ such that $\mu(\{0\})=0$, and suppose that  \begin{equation*}
 \int_{(0, 1]}\frac1\lambda \mu(d\lambda)<\infty.
\end{equation*} Further, suppose that $u_1,\ldots, u_n:\R\to\R\cup\{-\infty\}$, $i=1,\ldots,n$, are increasing and strictly increasing and differentiable on the interior of the respective domains $\operatorname{int}\operatorname{dom}u_i$, $i=1,\ldots,n$. If the allocation $(Z_1,\ldots, Z_n)\in \mathbb{A}_n(X)$ of $X\in L^1$ satisfies Borch's condition \eqref{eq:Borch2} and $Z_i=f_i(X)$ for some strictly increasing functions $f_i:\R\to \R$, $i=1,\ldots, n$, such that $\sum_{i=1}^nf_i=\operatorname{id}$, then $(Z_1,\ldots, Z_n)$ is Pareto optimal.
\end{corollary}

\begin{example}
Consider the exponential utility functions $u_i(x)=1-\exp(-\gamma_ix)$, $\gamma_i>0$, $i=1,\ldots,n$. 
Then Borch's condition \eqref{eq:Borch2} reads
$$\lambda_1 \gamma_1\exp(-\gamma_1 Z_1)= \ldots = \lambda_n\gamma_n\exp(-\gamma_n Z_n)$$
for constants $\lambda_i>0$, $i=1,\ldots, n$. 
This condition is satisfied if and only if $Z_i=\tfrac{1}{\gamma_i}aX+b_i$, $i=1,\ldots,n$, for constants $a,b_1,\ldots b_n$. 
The side constraint $X=\sum_{j=1}^n Z_j$ implies $\sum_{j=1}^n\tfrac{1}{\gamma_j}a=1$ and $\sum_{j=1}^n b_j=0$. 
Hence, any allocation
$$Z_i=\frac{1}{\gamma_i}\left(\sum_{j=1}^n\frac{1}{\gamma_j}\right)^{-1} X+b_i,\quad i=1,\ldots,n,$$ 
is Pareto optimal whenever $\Ucal_i=\Ucal_\mu$, $i=1,\ldots,n$, for $\mu\in \Mcal((0,1])$ such that $\mu(\{0\})=0$ and $\int_{(0, 1]}\frac1\lambda \mu(d\lambda)<\infty$. 
\end{example}

%\end{comment}

\end{appendix}

\end{document}